\documentclass[preprint]{aastex}

\usepackage{hyperref} 
\usepackage{amsmath}
\usepackage{lscape}

\usepackage{graphicx}
\usepackage{xspace}
\graphicspath{{./figures/}}
\usepackage{natbib}

\newcommand{\Fermi}{\emph{Fermi}\xspace}
\newcommand{\Swift}{\emph{Swift}\xspace}
\newcommand{\GRB}{{GRB~110731A}\xspace}
\newcommand{\Tz}[1]{{$#1$~s}}
\newcommand{\xspec}{\emph{XSPEC}\xspace}
\newcommand{\rmfit}{\emph{rmfit}\xspace}
\newcommand{\RA}[3]{\mbox{RA}={#1}$^{{\rm h}}${#2}$^{{\rm m}}${#3}$^{{\rm s}}$}
\newcommand{\decl}[3]{\mbox{DEC}={#1}$^{\circ}${#2}\arcmin{#3}\arcsec}

\newcommand{\cm}[1]{~cm$^{#1}$}

\newcommand{\cts}{~cts\,s$^{-1}$}
\newcommand{\e}[1]{10$^{#1}$}

\newcommand{\nh}{N$_{\rm H}$}
\newcommand{\be}{\begin{equation}}
\newcommand{\ee}{\end{equation}}
\newcommand{\ba}{\begin{eqnarray}}
\newcommand{\ea}{\end{eqnarray}}
\def\eps{\epsilon}

\begin{document}

\title{Multiwavelength observations of GRB~110731A: GeV emission from
onset to afterglow}

\author{
M.~Ackermann\altaffilmark{2}, 
M.~Ajello\altaffilmark{3}, 
K.~Asano\altaffilmark{4}, 
L.~Baldini\altaffilmark{5}, 
G.~Barbiellini\altaffilmark{6,7}, 
M.~G.~Baring\altaffilmark{8}, 
D.~Bastieri\altaffilmark{9,10}, 
R.~Bellazzini\altaffilmark{5}, 
R.~D.~Blandford\altaffilmark{3}, 
E.~Bonamente\altaffilmark{12,13}
% ...
}

\author{
S.~R.~Oates\altaffilmark{72},
T.~N.~Ukwatta\altaffilmark{33,75}, 
O.~Littlejohns\altaffilmark{76}
}

% ...
\altaffiltext{2}{Deutsches Elektronen Synchrotron DESY, D-15738 Zeuthen, Germany}
\altaffiltext{3}{W. W. Hansen Experimental Physics Laboratory, Kavli Institute for Particle Astrophysics and Cosmology, Department of Physics and SLAC National Accelerator Laboratory, Stanford University, Stanford, CA 94305, USA}
\altaffiltext{4}{Interactive Research Center of Science, Tokyo Institute of Technology, Meguro City, Tokyo 152-8551, Japan}
\altaffiltext{5}{Istituto Nazionale di Fisica Nucleare, Sezione di Pisa, I-56127 Pisa, Italy}
\altaffiltext{6}{Istituto Nazionale di Fisica Nucleare, Sezione di Trieste, I-34127 Trieste, Italy}
\altaffiltext{7}{Dipartimento di Fisica, Universit\`a di Trieste, I-34127 Trieste, Italy}
\altaffiltext{8}{Rice University, Department of Physics and Astronomy, MS-108, P. O. Box 1892, Houston, TX 77251, USA}
\altaffiltext{9}{Istituto Nazionale di Fisica Nucleare, Sezione di Padova, I-35131 Padova, Italy}
\altaffiltext{10}{Dipartimento di Fisica e Astronomia "G. Galilei", Universit\`a di Padova, I-35131 Padova, Italy}
% ...
\altaffiltext{12}{Istituto Nazionale di Fisica Nucleare, Sezione di Perugia, I-06123 Perugia, Italy}
\altaffiltext{13}{Dipartimento di Fisica, Universit\`a degli Studi di Perugia, I-06123 Perugia, Italy}
% ...
\altaffiltext{33}{NASA Goddard Space Flight Center, Greenbelt, MD 20771, USA}
% ...
\altaffiltext{72}{Mullard Space Science Laboratory, University College London, Holmbury St. Mary, Dorking, Surrey, RH5 6NT, UK}
% ...
\altaffiltext{75}{Michigan State University, East Lansing, MI 48824, USA}
\altaffiltext{76}{Department of Physics and Astronomy, University of Leicester, Leicester, LE1 7RH, UK}

\begin{abstract}
We report on the multiwavelength observations of the bright, long gamma--ray burst \GRB, by the \Fermi\ and \Swift\ observatories, and by the MOA and GROND optical telescopes. The analysis of the prompt phase reveals that \GRB shares many features with bright Large Area Telescope bursts observed by \Fermi during the first 3 years on-orbit: a light curve with short time variability across the whole energy range during the prompt phase, delayed onset of the emission above 100~MeV, extra power law component and temporally extended high--energy emission. In addition, this the first GRB for which simultaneous GeV, X--ray, and optical data are available over multiple epochs beginning just after the trigger time and extending for more than 800~s, allowing temporal and spectral analysis in different epochs that favor emission from the forward shock in a wind--type medium. The observed temporally extended GeV emission is most likely part of the high--energy end of the afterglow emission. Both the single--zone pair transparency constraint for the prompt signal, and the spectral and temporal analysis of the forward shock afterglow emission, independently lead to an estimate of the bulk Lorentz factor of the jet $\Gamma\sim$~500~--~550.

\end{abstract}




\section{Introduction}\label{intro}

Gamma--ray Bursts (GRBs) are the most powerful explosions in the Universe, initially releasing most of their energy in X--ray and gamma--ray on timescales lasting from a few seconds to a few minutes.  Highly variable light curves across the energy bands in this prompt emission phase suggest an active central engine that drives the highly--collimated GRB jet. The prompt emission is thought to be emitted by internal shocks, which are produced when shells of material collide within the jet. The  variability within this emission, as observed in the prompt light curves, is thought to be due to intermittent central engine activity~\citep{1994ApJ...430L..93R}.

However, the details of the emission mechanism which can explain the efficiency of the internal shocks are not understood. Fainter and longer-lived emission following the prompt phase, called the GRB afterglow~\citep{1997ApJ...476..232M, 1998ApJ...497L..17S}, has been observed at lower energies prior to \Fermi, ranging from X--ray to optical and radio wavelengths.  The first observations by the \Fermi observatory of delayed and long-lived GeV emission relative to the prompt MeV emission~\citep{2009ApJ...706L.138A, 2009Sci...323.1688A} have led to speculation~\citep{2009MNRAS.400L..75K,MNR:MNR17274,2010MNRAS.403..926G, 2010ApJ...709L.146D, 0004-637X-720-2-1008, 2010ApJ...724L.109R} that the afterglow component may also include a significant amount of gamma--ray emission from the high energy tail of the synchrotron radiation of the external forward shock. A contribution from the fterglow to the gamma--ray flux at such early times would indicate a significantly earlier the onset of the interaction between the GRB blast wave and the circum--burst medium~\citep{1976PhFl...19.1130B}, which is thought to be the source of the afterglow emission. The theoretical models of the underlying afterglow emission~\citep{2002ApJ...568..820G} also constrain the physical parameters of the external shock such as the jet energy, bulk Lorentz factor, emission efficiency and nature of the surrounding medium when optical--to--GeV data are fit simultaneously in this framework.

The \Fermi observatory hosts two instruments, the Large Area Telescope (LAT) which covers an energy range from 20 MeV to up to more than 300 GeV~\citep{2009ApJ...697.1071A} and the Gamma--ray Burst Monitor (GBM) which is sensitive at lower energies, from 8 keV to 40 MeV~\citep{2009ApJ...702..791M}. Together, the LAT and GBM are capable of measuring the spectral parameters of GRBs across seven decades in energy. During the first 3 years on-orbit, the \Fermi LAT has detected more than 30 GRBs with high significance, but only four of these have benefited from simultaneous detections with the \Swift Burst Alert Telescope (BAT), allowing for prompt follow--up with the narrow--field X--Ray Telescope (XRT) and Ultraviolet Optical Telescope (UVOT) on \Swift~\citep{Gehrels04}. The first such burst was GRB~090510, a very bright short GRB for which the afterglow was observed contemporaneously by \Fermi LAT and by \Swift\ XRT and UVOT starting just 100 s after the burst trigger~\citep{2010ApJ...709L.146D}.  Multiwavelength data from GRB~090510 seem to favor an afterglow interpretation as their origin~\citep{2010ApJ...709L.146D, 0004-637X-720-2-1008, 2010ApJ...724L.109R}.  The second such burst was GRB~100728A, a very long burst for which \Fermi revealed significant temporally extended emission out to 850~s post trigger, and \Swift\ detected a series of strong X--ray flares in the XRT light curves.  An internal shock scenario seems to reproduce well both the prompt emission and the later X--ray flares as well as temporally extended high--energy emission from GRB~100728A~\citep{2011ApJ...734L..27A}. The third was GRB~110625A~\citep{2012ApJ...754..117T} which is a burst that triggered both BAT and GBM but which was outside of the LAT field--of--view during the prompt phase. However, thanks to the \Fermi Autonomous Repoint Request (ARR) there was a joint XRT and LAT detection of the temporally extended emission.

In this paper we report on the analysis of the bright and long \GRB, the fourth burst to benefit from joint \Fermi and \Swift\ observations, and one with the most comprehensive multiwavelength data of any LAT--detected GRB to date.  Fortuitously, the burst was within the LAT field of view at the trigger time and caused an ARR.  BAT also triggered on the burst and \Swift was immediately repointed for XRT and UVOT observations. In addition, the burst was also observed by ground--based observatories, including the Microlensing Observations in Astrophysics (MOA) telescope at early and intermediate times, and the Gamma--ray Burst Optical/Near-Infrared Detector (GROND) at late times. Multiwavelength observations hence cover both the prompt phase and the temporally extended emission of \GRB, from optical to GeV energies.

The paper is organized as follows: Section~\ref{section:obs} includes a comprehensive description of the observations made by the various instruments that detected the burst's prompt and temporally extended emission, Section~\ref{section:analysis} gives the details of the data reduction and analysis for both the prompt emission and temporally extended emission for all telescopes, Section~\ref{sec:results} reports the results of the prompt and temporally extended emission data analysis in the multiwavelength context and Section~\ref{sec:disc} provides the discussion and interpretation of both the prompt and temporally extended emission results, and finally we draw our conclusions in Section~\ref{sec:conclusions}.

Throughout the paper, times $t=T-T_0$ are given relative to the GBM time of trigger $T_0$, and the afterglow convention for the energy flux $F_{\nu,{\rm t}} \propto \nu^{-\beta}  t^{-\alpha}$  has been followed, where the energy index $\beta$ is related to the differential photon index $\Gamma=\beta+1$. The phenomenology of the burst is presented in the reference frame of the observer, unless otherwise stated. All the quoted errors are given at the 68\% confidence level for one parameter of interest.

\section{Observations}
\label{section:obs}
On 2011 July 31 at 11:09:29.94 (UT), GBM triggered on \GRB which, due to the high peak flux of this burst, caused an ARR. \GRB was already well within the LAT field of view, being only $\sim3.3^\circ$ off axis so that the repointing had little impact on the prompt emission phase observations. The maneuver placed the spacecraft in pointing mode for 2.5 hours after the burst, allowing continuous LAT observation of the burst from the initial time of trigger until the first \Fermi passage into the South Atlantic Anomaly (SAA) at \Tz{1400}. The ARR continued for another 90 minutes after \Fermi had exited the SAA at \Tz{3150}, with the burst being well within the LAT field of view from \Tz{4000} to \Tz{7400}.  No significant signal from \GRB was found at these late times.

The best LAT localization of \GRB is  \RA{18}{41}{00}, \decl{$-$28}{31}{00} (J2000), with a 68\% confidence error radius of 0.2$\degr$~\citep{2009ApJ...707..580A}

GRB110731A triggered the \Swift BAT at 11:09:30.45 UT.
\Swift~slewed immediately to the burst, and its narrow--field instruments, XRT and UVOT, began observations 56~s after the BAT trigger.  An accurate afterglow position was rapidly determined by the UVOT as \RA{18}{42}{00.99}, \decl{$-$28}{32}{13.8} (J2000, \citealt{swiftgcnreport}), with an error radius of 0.5~arcsec (90\% confidence).

XRT observations started while the spacecraft was settling at the end of the initial slew. The XRT  began collecting data in  Window Timing (WT) mode, 
as the source was bright ($\sim$100\cts), and automatically switched 
to Photon Counting (PC) mode when the count rate from the source decreased to $<2$\cts.
Follow--up X--ray observations occurred during the following 24 days for a total net exposure of 600~s in WT mode and 75~ks in PC mode.

The UVOT took a short exposure with the {\em v} filter during the settling phase.
This exposure was followed by a `finding chart' exposure with the \emph{White} filter lasting 147~s.
UVOT then began its usual procedure of cycling through its 3
visible filters (\textit{v}, \textit{b}, and \textit{u}) and 3 \textit{UV} filters (\textit{uvw1}, \textit{uvm2}, and \textit{uvw2})~\citep{2008MNRAS.383..627P,2010MNRAS.406.1687B}. The optical afterglow was detected in the \emph{White}, \textit{u}, \textit{b}, and \textit{v} filters, but not in the \textit{UV}  filters. The lack of detection in the \textit{UV} filters is consistent with the measured redshift of $z=2.83$~\citep{gcn12225}.

MOA observations began 3.3~min after the \Swift~trigger for \GRB \citep[GCN~12242, ][]{gcn12242}.
Using a 61~cm Boller \& Chivens telescope at the Mt. John University Observatory in New Zealand, I and V band images with 60 s exposures followed by 120~s exposures until 12:56 UT (105 min after the trigger) were obtained. The total numbers of I and V images are 39 and 35, respectively; however due to the difficulty of photometry in the crowded field, we used only the 30 $I$ and 19 $V$ data points reported in Tab.~\ref{Tab:MOA-all}.

After a weather--induced delay, the seven--color imager GROND~\citep{2008PASP..120..405G} mounted on the 2.2~m MPG/ESO telescope at La Silla Observatory, Chile,  observed \GRB at a mean time of 2.74 days after the trigger. Two 30--minutes observation blocks were obtained
which yielded an integration time of 4500~s in $g^\prime r^\prime i^\prime z^\prime$ and 3600~s in $JHK$. The mean seeing during the observations was 1\farcs3.

\section{Data Analysis}
\label{section:analysis}

\subsection{\Fermi}
\label{section:prompt}

For the time--integrated and time--resolved spectral analysis of the prompt phase, both GBM and LAT data were used. The \Fermi LAT and GBM data may be retrieved from the \Fermi Science Support Center archives~\footnote{http://heasarc.gsfc.nasa.gov/FTP/fermi/data/gbm/triggers/},~\footnote{http://fermi.gsfc.nasa.gov/ssc/data/access/}.

The GBM detectors were selected in the same fashion as outlined in~\citep{2009ApJ...707..580A, 2011A&A...528A..15G, 2012ApJS..199...19G}: we used the sodium iodide (NaI) detectors 0 and 3, and bismuth germanate (BGO) detector 0. We also used \emph{Time Tagged Events} TTE data~\citep{2009ApJ...702..791M} for our spectral analysis with a temporal resolution of 64~ms, in the 8~keV to 40~MeV energy range, excluding the range around the NaI K--edge at 33.17~keV.

For the LAT, we first extracted `P7TRANSIENT'--class data from a circular region centered on the burst position with energy--dependent radius equal to a 95\% containment of the point--spread function (PSF), see~\cite{2009ApJ...707..580A} for details. To greatly reduce the numbers of gamma rays from the Earth limb we selected events with zenith angles less than 100$\degr$. We then followed the procedure described in~\cite{2009ApJ...707..580A} to estimate the residual background and considered front-- and back--converting events separately~\citep{2009ApJ...697.1071A}. The results presented here were obtained using the \Fermi \emph{ScienceTools--v9r25p1} and \emph{P7TRANSIENT\_V6} instrument response functions (IRFs).  

Besides the `P7TRANSIENT' data, we also extracted events using the so--called \emph{LAT low energy} (LLE, E$\ge10$ MeV) events selection criteria~\citep{LLE}: This selection keeps events that pass the \texttt{GAMMA} filter, have a reconstructed track in the tracker pointing to a sky location that is roughly compatible with the GRB position~\citep{2009ApJ...697.1071A}. By retaining very low energy events, this selection provides high statistics light curves that are useful for temporal analysis, see Section~\ref{sec:promptresults}. We did not use LLE data  for the spectral analysis since LLE and `P7TRANSIENT' data gave consistent results.

The LAT/GBM joint spectral fits were performed with the software package \xspec version 12.7.0e \citep{1996XspecProc}, and cross--checked with \rmfit version 4.0rc1~\footnote{{\it rmfit} for GBM and LAT analysis was developed by the GBM Team and is publicly available at fermi.gsfc.nasa.gov/ssc/data/analysis/}~\citep{2006ApJS..166..298K,2009ApJ...707..580A}. GBM Response Matrices v1.8 were used with both tools. For the fitting procedure, the PG-statistic~\citep{2011XspecBook} (S in the following) was used within \xspec, while the Castor statistic~\citep{2011ApJ...729..114A} was used for \rmfit.

For each time interval of interest, we compared the fit of several models that are listed below (differential fluxes are in ph/cm$^2$/keV/s): 
\begin{itemize}
\item[(i)]{a power law with an exponential cutoff (Comptonized model, hereafter COMP), whose differential photon flux is described by the equation:
\begin{equation}
\text{COMP}(E) = A \left(\frac{E}{E_{ref}}\right)^{\Gamma_\alpha}\exp\left(-\frac{E}{E_{0}}\right),
\end{equation}
where $A$ is the normalization amplitude, $\Gamma_\alpha$ is the photon index, $E_{ref}$ is the reference energy fixed at 1~MeV and $E_0$ is the cutoff energy.}
\item[(ii)]{a Band function, hereafter BAND, as defined in~\citep{1993ApJ...413..281B}, where $A$ is the normalization amplitude, $\Gamma_\alpha$ and $\Gamma_\beta$ are the low and high energy power law indices and $E_0$ is the cutoff energy:}
\begin{equation}
\text{BAND}(E) = 
\begin{cases}	
A \left(\frac{E}{100\,{\rm keV}}\right) ^{\Gamma_\alpha} \exp(-\frac{E}{E_0}) & \text{$E \le (\Gamma_\alpha - \Gamma_\beta) E_0$} \\
A  \left[ \frac{(\Gamma_\alpha-\Gamma_\beta)E_0}{100\,{\rm keV}}\right]^{\Gamma_\alpha-\Gamma_\beta} \left(\frac{E}{100\,{\rm keV}}\right)^{\Gamma_\beta} \exp(\Gamma_\beta-\Gamma_\alpha) & \text{$E > (\Gamma_\alpha - \Gamma_\beta) E_0$}
\end{cases}
\end{equation}
\item[(iii)] {a COMP model plus a power law (COMP+PL):
\begin{equation}
  \text{COMP+PL}(E)=\text{COMP}(E)+B\times \left(\frac{E}{100\,{\rm keV}}\right)^{\Gamma_\gamma}
\end{equation}
 where $B$ is the normalization and $\Gamma_\gamma$ the power law photon index.}
\item[(iv)]{a BAND model plus a power law (BAND+PL):
\begin{equation}
  \text{BAND+PL}(E)=\text{BAND}(E)+B\times \left(\frac{E}{100\,{\rm keV}}\right)^{\Gamma_\gamma}
\end{equation}
}
\item[(v)]{a model with a sum of two independent Comptonized components (COMP+COMP) that in addition to the extra PL component provides also a second high--energy cutoff.}
\end{itemize}
We remind the reader that the commonly referenced peak energy $E_{\text{peak}}$ of the $\nu F_{\nu}$ spectrum can be computed as:
\begin{equation}
E_{\text{peak}} = (2+\Gamma_\alpha)E_0\;\;\;\;\;\;\text{in the COMP model.}
\end{equation}

The COMP (i) and BAND (ii) models reported above were historically found to give very good empirical descriptions of GRB spectra~\citep{1993ApJ...413..281B,2006ApJS..166..298K}, with the COMP model having one less free parameter and providing a sharper decrease at high energies (usually in the $\sim$MeV range) than the BAND model.
In order to check whether a different spectral component (meaning a different physical process and potentially a different location of the emission) could be responsible for the GeV emission observed in the LAT data, we added a power law component to models (i) and (ii), resulting the new models COMP+PL (iii) and BAND+PL (iv). Additionally, as some theoretical models predict that the high--energy GeV emission must be cut off owing to gamma-gamma pair production opacity considerations, see Section~\ref{sec:disc:prompt}, we tested this hypothesis by considering the COMP+COMP model (v), that matches these characteristics while keeping a reasonable number of free parameters.

For each time interval we determined the best--fit model that provides a good description of the data with a minimal set of parameters, following the method described in Section 4.1 of~\cite{2011ApJ...729..114A}. In particular, we used the likelihood ratio test (LRT) to derive the significance of the improvement of the fit when comparing a simpler model (the null hypothesis) with a more complex model (the alternative hypothesis). Using the PG-stat value, $S$, of the statistic and defining $\Delta S$ as the difference between the values of $S$ obtained with the two models, the LRT gives the probability $P(\Delta S)$ that the observed $\Delta S$ has been obtained because of statistical fluctuations, on the assumption that the null hypothesis is the true model. Thus, if $P(\Delta S)$ is low the alternative hypothesis is to be preferred. We sampled the distribution of $\Delta S$ via 10 millions Monte Carlo realizations of the burst spectrum with \xspec. With such large statistics, we were able to compare pairs of models for probabilities down to $P(\Delta S) \sim 1 \times 10^{-7}$. Such simulations cannot account for systematic effects, for example due to the uncertainties in the responses of the instruments. Although a $10^{-4}$ probability would be formally very significant, we adopted a conservative threshold $P_{\text{th}} = 1\times 10^{-5}$, and we preferred the alternative models over the null hypothesis if $P(\Delta S) < P_{\text{th}}$.

Since the count fluence of this burst is very high, we used the effective area correction factors~\citep{2009ApJ...706L.138A} to account for possible calibration issues between the different detectors. We used the time--integrated spectrum (interval P1, see Section~\ref{sec:spectrum}) and the BAND model to determine these factors, which were 0.85 for the BGO detectors and 0.96 for the NaI detectors, using the LAT spectrum as a reference. We applied these effective area corrections throughout the spectral analysis.

We explore the temporally extended emission from the burst in the LAT data running an unbinned likelihood analysis with the \Fermi \emph{ScienceTools--v9r25p1} and \emph{P7TRANSIENT\_V6} IRFs.  Again, we estimated the residual particle background following the procedure described in~\cite{2009ApJ...707..580A} for each sub--data selection. The model prepared for the likelihood analysis was then composed of a single power law component for the burst itself, the particle background template and the template model `gal\_2yearp7v6\_v0.fits' for the Galactic diffuse emission~\footnote{\Fermi background models are available from the FSSC web site at \href{http://fermi.gsfc.nasa.gov/ssc/data/access/lat/BackgroundModels.html}{http://fermi.gsfc.nasa.gov/ssc/data/access/lat/BackgroundModels.html}} for which the normalization was held fixed during the fit.

In order to test the significance of the detection/light curve of the GRB at a particular time, without being dependent upon a subjective choice of time interval, we determined interval boundaries through the use of an algorithm that extends the time intervals until the likelihood test statistic is 18 in each bin.  If a time interval exceeds the good time interval, an upper limit is reported and the computation steps to the next good time interval. The minimum number of events required in each interval is 8, to guarantee a reasonable number of degrees of freedom in the fit. Note that this method also optimizes the detection probability and and provides the maximum number of time intervals on which the analysis can be run (more details may be found in~\cite{2012A&A...544A...6L}).

In order to fit the spectral energy densities (SEDs) using data from the different instruments (see Section~\ref{sec:sed} below), it was necessary to combine and adjust the LAT time intervals to have sufficient counts per bin and to match the observed features of the XRT light curve (see Section~\ref{sec:resultsafterglow} below). We defined the new time intervals as follows: [8.3, 11.5]~s (I, hereafter), [11.5, 55.0]~s (II), [55.0, 227.0]~s (III) and [227.0, 853.9]~s (IV). 
We fit the LAT again over these time intervals using a slightly different background model in which the particle background and the Galactic diffuse emission are both estimated using the procedure described in~\citet{2009ApJ...707..580A} so that we could extract separately the signal and background data to be used as input to \xspec for the joint fits of the SEDs.

\subsection{\Swift}
\label{sec:swift}
We retrieved the \Swift~data from the HEASARC archive\footnote{
http://heasarc.gsfc.nasa.gov/docs/swift/archive/}
and processed them with the standard \Swift~analysis software (v3.8)
included in the NASA's HEASARC software (HEASOFT, ver.~6.11) and the relevant calibration files.

We extracted BAT mask--weighted light curves and spectra in the nominal 15--150 keV 
energy range following the standard procedure~\citep{2008ApJS..175..179S}. 
The BAT data were not used for the spectral analysis of the prompt phase because the GBM data alone constrain $\Gamma_\alpha$ very well and because the cross--calibration between BAT and GBM is still not well understood and the subject is beyond the scope of this paper.

We extracted the XRT light curves and spectra in the nominal 0.3--10 keV energy 
range by applying standard screening criteria. All the XRT data products 
presented here are background subtracted and corrected for 
PSF losses, vignetting effects and exposure variations~\citep[see][]{2007A&A...469..379E,2009MNRAS.397.1177E}.

UVOT photometric measurements were complicated by the crowded field.
We obtained the source count rates from a circular source extraction region 
with a radius of 5\arcsec~\citep{2008MNRAS.383..627P,2010MNRAS.406.1687B}. 
We estimated the background from nearby circular regions with radii of 20\arcsec,
whithin which field sources were masked out. We also used \Swift~late--time observations to estimate the residual contribution of nearby objects, and to refine the afterglow photometry. For this reason, the last data point used in the light curve is at \Tz{1000}. In order to better constrain the optical temporal decay we created a single light curve (see Fig.~\ref{Fig:mwllc}) from all the UVOT filters by renormalizing each light curve to the \emph{v}--band~\citep{2009MNRAS.395..490O}, using flux conversion factors from~\cite{2011AIPC.1358..373B}.

\subsection{MOA}
We performed the MOA data analysis using aperture photometry via the \emph{SExtractor} package~\citep{1996A&AS..117..393B} to estimate the instrumental magnitudes of the objects in each image.  We then compared these values to the late time GROND data of the same field for zero point determination of each of the MOA images.
Since the MOA data were obtained with Bessell's {\em I} and {\em V} band filters, which have transmission curves similar to those of Johnson--Cousin's {\em I} and {\em V} band filters in combination with the CCD quantum efficiency curve, and GROND uses the {\em g'z'r'i'} filter system, we used the conversion table in \citet{2002AJ....123.2121S} to obtain a GROND equivalent {\em I} and {\em V} band magnitudes with which to perform this comparison. In order to reduce the systematic error in the zero point determination, we selected only the top 30th percentile of brightest stars in the MOA images.  Finally, we produced light curves for the same bright stellar objects in each of the MOA images for a final relative calibration.  The resulting median $\Delta m$ variations in these stellar light curves were used as additional corrections to the afterglow light curve that account for any errors in the zero point determination of the individual images. We estimated the systematic error in the zero point calibration of each image by measuring the standard deviation of the difference between the MOA and GROND equivalent {\em I} and {\em V} band magnitudes in each image. We then summed the statistical errors of the measured fluxes as returned by \emph{SExtractor} in quadrature with this systematic error to obtain the error estimate in the final MOA flux density data. MOA results are given in Tab.~\ref{Tab:MOA-all}, and as fluxes reported on the multiwavelength light curve shown in Fig.~\ref{Fig:mwllc}.

\subsection{GROND}
We analyzed the GROND data analysis with a custom routine as described in~\cite{2008ApJ...685..376K} and~\cite{2008AIPC.1000..227Y}, using \emph{SExtractor}  for background subtraction, and masking out bright sources. At the position of the afterglow, a faint source was visible in $r^\prime i^\prime z^\prime$, but photometry was hindered by multiple nearby stars in the crowded field. Therefore, we obtained obtained measurements for a second epoch on 2011 September 25, with identical exposure time, under improved conditions, to create a template image for image subtraction. We performed the image subtraction using \emph{HOTPANTS}\footnote{http://www.astro.washington.edu/users/becker/hotpants.html}. We used 60 different parameter settings in determining the Gaussian PSF kernel for the subtraction routine for each band, and chose the combination of parameters that resulted in the best subtraction of nearby stars near the afterglow position, as measured by the noise in the residual image at the afterglow position. In the residual image, the afterglow is strongly detected in $r^\prime$, still well--detected in $i^\prime z^\prime$, only faintly detected in $g^\prime JH$ and undetected in $K$. We calibrated the magnitudes of stars in the field against an SDSS standard star field at similar RA observed just before the first epoch observations under photometric conditions. We performed the photometry using seeing--matched aperture photometry on the subtracted images with MIDAS\footnote{http://www.eso.org/sci/software/esomidas/}. We estimated the errors on the fluxes as the sum in quadrature of the calibration error, the statistical error of the detection, and the noise error of the image subtraction as determined by \emph{HOTPANTS}. GROND results are given in Tab.~\ref{Tab:grond}, and as fluxes reported on the multiwavelength light curve graph shown in Fig.~\ref{Fig:mwllc}.

\section{Results}
\label{sec:results}

\subsection{Prompt phase}
\label{sec:promptresults}

\subsubsection{Light curves and Timing results}\label{subsection:lightcurves}
\begin{figure}[htbp]
  \centering
  \includegraphics[scale=0.60]{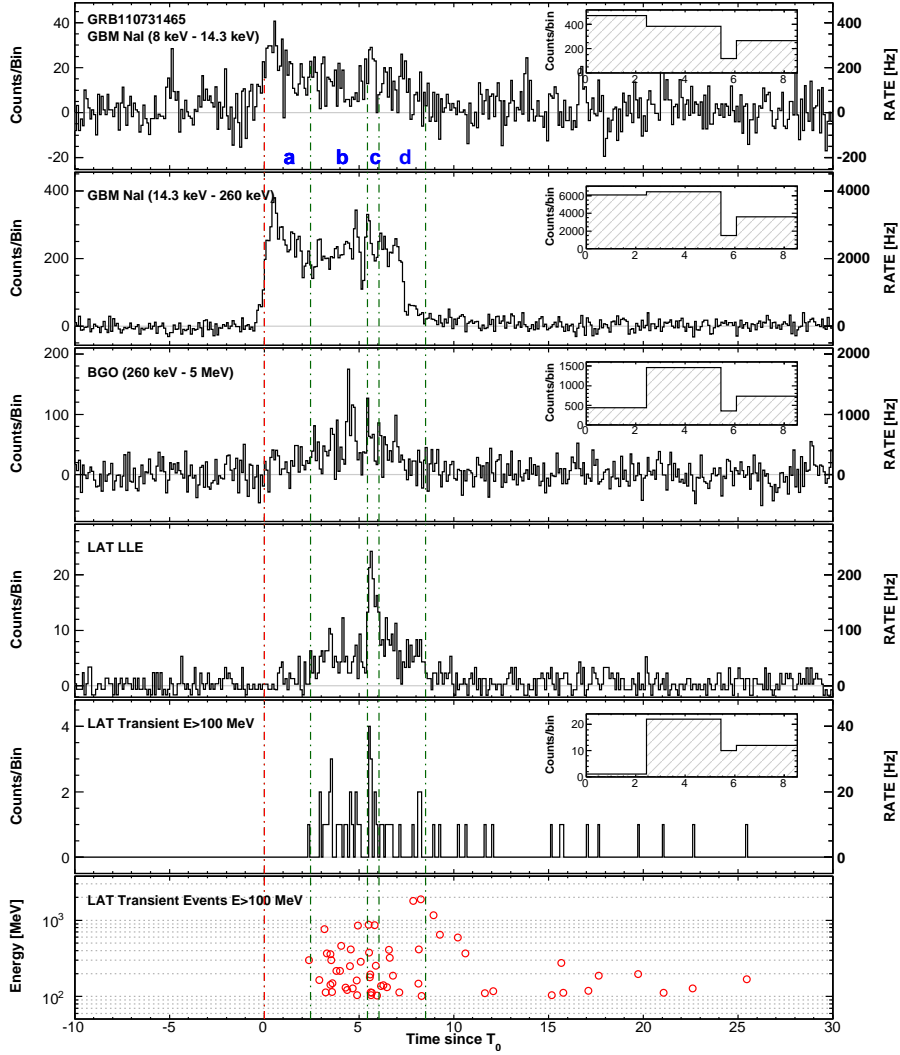}
  \caption{GBM and LAT light curves for the gamma--ray emission of \GRB. 
The data from the GBM NaI detectors were divided into soft (8--14.3~keV) 
and hard (14.3--260~keV) bands to reveal similarities between the light curves 
at the lowest energies and for the LAT data.
The first four light curves are background-subtracted and have 0.1~s time binning.
The fourth panel shows the LAT LLE light curve~\citep{LLE}.
The fifth panels shows LAT `P7TRANSIENT'--class events light curves for energies $>$ 100~MeV, with 0.5\,s time binning. The sixth panel shows the energy and arrival time of LAT `P7TRANSIENT'--class events above 100~MeV
The vertical lines indicate the boundaries of the intervals \emph{a, b, c, d},
used for the time--resolved spectral analysis for boundaries [0.00, 2.44, 5.44, 6.06, 8.52]~s. 
The insets show the counts for each data set, 
binned using these intervals, to illustrate the numbers of counts considered 
in each spectral fit.}
  \label{Fig:promptlc}
\end{figure}

In Fig.~\ref{Fig:promptlc}, we show the GBM and LAT light
curves of the GRB prompt emission phase in several energy bands, from 8~keV to above 1~GeV.  The light curves show a complex multi--peaked structure and have two interesting features: (i) the LAT emission at $>10$ MeV is slightly delayed ($\sim2.5$ s) with respect to the GBM light curves, (ii) a peak with high count rate is also present at \Tz{5.5} in the LAT data that is also present in the NaI and BGO light curves.

Detailed analysis of the GBM data results in a  $T_{90,\rm{GBM}}$ duration~\citep{1993ApJ...413L.101K} in the 50~keV to 300~keV energy range of $7.3\pm0.3$~s, with a start time defined by $T_{05,\rm{GBM}}=0.25\pm0.1$~s and an end time at $T_{95,\rm{GBM}}=7.6\pm0.3$~s. A similar detection and duration analysis of the LLE light curve ($>10$~MeV) demonstrated that the LAT prompt phase detection starts later than observed in the GBM, at $T_{05,\rm{LLE}}=2.5^{+0.3}_{-0.6}$~s, and lasts longer with $T_{90,\rm{LAT}}=14.3^{-2.6}_{+17.0}$~s. These results for the LLE event selection are consistent with the findings for `P7TRANSIENT'--class events ($>100$~MeV) using an estimation of the total background following the procedure described in detail in~\citet{2009ApJ...707..580A}. Indeed, for the `P7TRANSIENT' events we found a comparable $T_{05,\rm{LAT}}=3^{+0.3}_{-0.2}$ s. This analysis also revealed temporally extended emission up to $T_{95,\rm{LAT}}=190^{+70}_{-170}$ s.

The LLE light curve was investigated using a Bayesian blocks algorithm~\citep{1998ApJ...504..405S, 2012arXiv1207.5578S} to determine intervals over which the photon arrival rate has no statistically significant variations. By requiring a large statistical significance for the rate variations (to be less sensitive to background noise) four time intervals were found at [0, 2.44, 5.44, 6.06, 8.52]~s and are highlighted in Fig.~\ref{Fig:promptlc}. The upper boundary of the first time interval is at \Tz{2.44} which is consistent with the previous estimate of the first signal significant detection in the LAT as reported above with other techniques. The four time intervals will be used in the next Section for the time--resolved spectral analysis.

The flux peak at \Tz{5.5} is observed across the whole energy spectrum so we examined it by calculating the cross--correlation functions (CCFs), as defined in~\cite{1988ApJ...333..646E}, between energy bands to quantify the simultaneity of the emission. We calculated the CCF between the low energy NaI and the LLE light curves, first considering the full prompt phase from $-1$~s to \Tz{8.52}: the CCF has a local maximum at $0.3\pm^{0.5}_{0.1}$ s, consistent with no lag. However, a second CCF maximum occurs at \Tz{5.2}, possibly due to overlapping pulses and making the lag measurement subject to systematic uncertainties. The CCF thus suggests that the peak is indeed statistically coincident in time across the whole energy range, although it is less prominent at the very lowest energies in the NaI detectors than at high energies (the LAT events above 100~MeV). 

We further characterized the variability of the emission by deriving a typical variability time scale for the burst emission following the light curve pulse deconvolution technique described in~\citet{2011arXiv1109.4064B}. We used the summed light curves from the 4 brightest NaI detectors, both the BGO detectors, as well as the LLE data. The light curves were sub--divided into various energy bands, when possible, to estimate the energy dependence of the full width at half maximum (FWHM) of the pulses. The median value of the FWHM of the pulses in the GBM NaI and BGO data over the $T_{90,\rm{GBM}}$ interval shows a weak dependence on energy, decreasing from $0.35\pm 0.02$~s at $\sim 18$~keV to $0.24\pm 0.01$~s at 2~MeV with an average value $0.28\pm 0.02$~s. The median value of the FWHM in the LLE data during the $T_{90,\rm{GBM}}$ interval is larger, $0.92\pm 0.15$~s at $\sim 17$~MeV. In a shorter time interval, [\Tz{3.0}, \Tz{7.6}], that excludes the $T_{05,\rm{LAT}}$ time, the median value for the FWHM is $0.43\pm 0.03$~s with a minimum of $0.147\pm 0.003$~s, in the 8~keV--1~MeV range.

For each LAT `P7TRANSIENT'--class event, we estimated the probability of its being associated with the GRB using the {\em gtsrcprob} \Fermi Science Tool. The probability computation takes into account the spectral and spatial distributions of all of the components in the source model, convolved with the response of the LAT as well as the exposure (all convolved with the effective PSF). The values of the model parameters are found via a maximum likelihood analysis~\citep{1996ApJ...461..396M}, and the probability of a particular event being attributed to a particular source component is proportional to the predicted counts density for the event by that component. The highest photon energy during the prompt phase is a 2.0 GeV event at \Tz{8.27}, having a probability of $10^{-6}$ to be associated with the background. During the temporally extended emission phase, we note that a 3.4 GeV event at \Tz{435.96} has a probability of $\sim10^{-3}$ to be associated with the background.

\subsubsection{Spectral analysis}
\label{sec:spectrum}
We analyzed the burst emission spectrum over a number of time intervals:
\begin{itemize}
\item[-] the 4 intervals $a,b,c,d$ which were determined by the Bayesian blocks analysis in Section~\ref{subsection:lightcurves}.
\item[-] interval P1 ($a+b+c+d$ or [0, \Tz{8.52}]) corresponds to the entire prompt emission phase.
\item[-] interval P2 ([\Tz{3.0}, \Tz{7.6}]) spans the time range $T_{05,\rm{LAT}}$ to $T_{95,\rm{GBM}}$ in which we observe the maximum flux in both instruments.
\end{itemize}

We first considered the time interval P1 for the time--integrated spectral studies.
We found that BAND is the preferred model. Both COMP+PL and BAND+PL provided only limited improvements in the fit. Comparing each one separately with BAND, we obtained a null hypothesis probability $P(\Delta S)\simeq 1 \times 10^{-4}$ for both. The best--fit parameters for BAND were $\Gamma_{\alpha}=-0.89^{+0.03}_{-0.03}$, $\Gamma_{\beta}=-2.32^{+0.03}_{-0.03}$, $E_{0}=324^{+27}_{-25}$~keV with a corresponding fluence in the 10 keV -- 10 GeV energy band of $F=(4.56 \pm 0.05)\times10^{-5}\; \rm{erg\;cm}^{-2}$, calculated in the rest frame.

\begin{figure}[htbp]
  \centering
  \includegraphics[width=\textwidth]{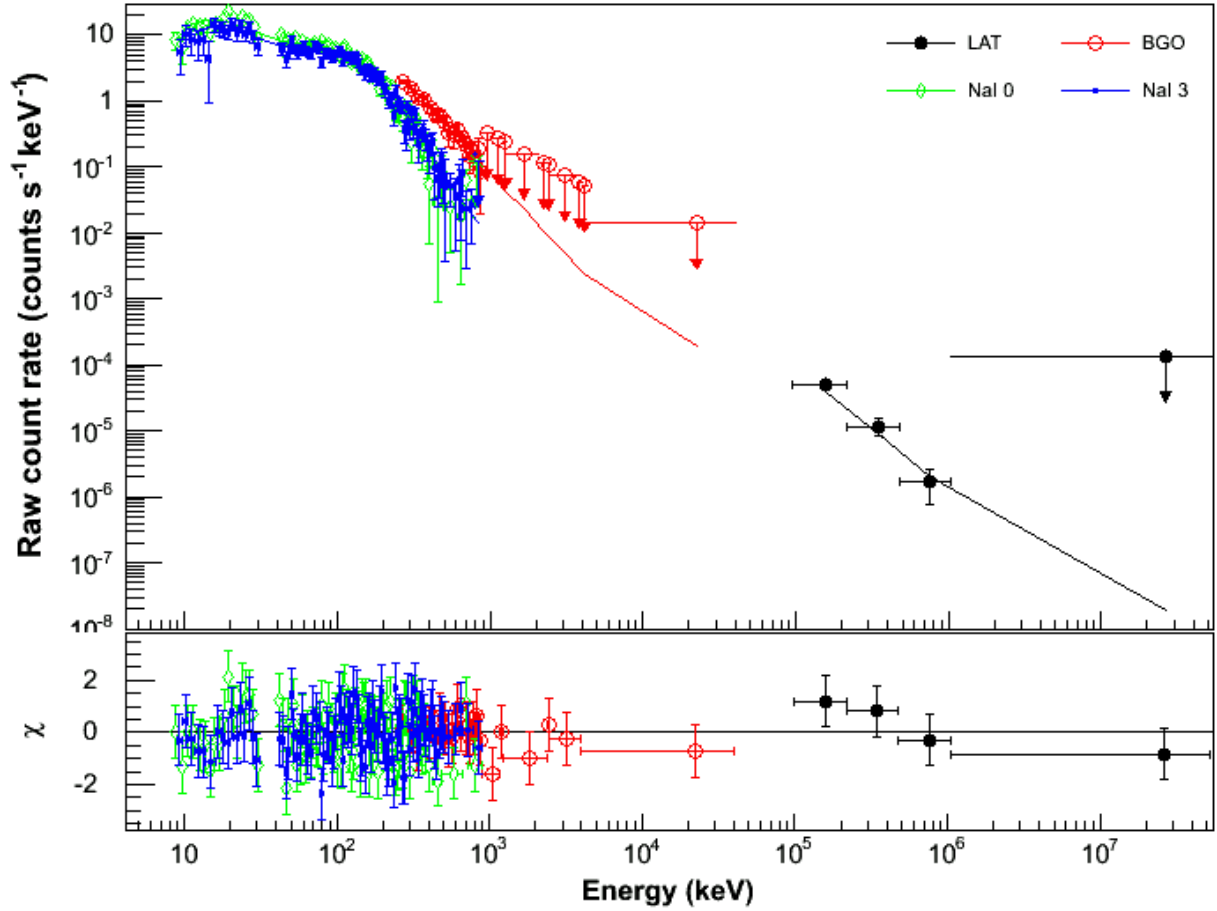}
  \caption{Joint spectral fitting of GBM and LAT data for the time interval P2, [\Tz{3.0}, \Tz{7.6}]. The top panel shows the count spectra (points) and best--fit BAND+PL model (lines). The lower panel shows the residuals.}
  \label{Fig:counts}
\end{figure}

To further investigate the significance of the additional power law component, we performed a time--integrated spectral analysis on interval P2. Similarly to interval P1, the BAND model fitted the data reasonably well. This time, however, BAND+PL gave a large improvement and smaller residuals with respect to BAND, with $\Delta S = 35.2$ and a corresponding null hypothesis probability of $P(\Delta S)< 1 \times 10^{-7}$. Thus, the power law component is required to account properly for the high--energy part of the spectrum. Moreover, the COMP+COMP model provided an even better fit for the same number of degrees of freedom, with $\Delta S = 15.5$ when compared to BAND+PL. Using the latter as null hypothesis we obtain $P(\Delta S) \simeq 3 \times 10^{-5}$. This latter result strongly suggests the presence of a cutoff in the high--energy part of the spectrum, although the null hypothesis probability is not formally below $P_{\text{th}}$.  A COMP+BAND or BAND+BAND model did not improve the fit over the COMP+COMP model.  Furthermore the power law slope of the high--energy BAND component could not be constrained due to limited statistics. The best--fit parameters of this time interval are given in Tab.~\ref{Tab:lat05gbmt95}, and the count spectrum corresponding to BAND+PL is shown in Fig.~\ref{Fig:counts}.

\begin{table*}[htbp]
\centering
\begin{tabular}{llllll}

\hline
Fitting model		&				&	BAND				&		BAND+PL			&	COMP+PL				&	COMP+COMP					\\
\hline\\[2pt]
$E_0$ [keV]		&				&	349$_{-28   }^{+31    }$	&	155$_{-13    }^{+20  }$	&	198.8$_{-18    }^{+21  }$	&	191.1$_{-18    }^{+21  }$		\\[2pt]
$\Gamma_\alpha$	&				&	$-0.74_{-0.04 }^{+0.04 }$	&	 0.03$_{-0.12 }^{+0.15 }$	& 	$-0.14_{-0.10 }^{+0.10 }$	&	 $-0.10_{-0.12 }^{+0.12 }$		\\[2pt]
$\Gamma_\beta$			&				&	$-2.31_{-0.03 }^{+0.03 }$	&	$-2.40_{-0.20  }^{+0.10  }$&	$-$					&	$-$						\\[2pt]

\hline \\[2pt]

\textit{extra component}	&				&						&						&						&							\\[2pt]
$\Gamma_\gamma$			&				&	$-$					&	$-1.96_{-0.05 }^{+0.09 }$	&	$-1.89_{-0.02 }^{+0.02 }$	&	$-1.79_{-0.03 }^{+0.03 }$		\\[2pt]
Cutoff Energy [MeV] &				&	$-$					&	$-$					&	$-$					&	390$_{-120    }^{+220  }$		\\
	
\hline\\[2pt]

Fluence [$10^{-5}\,\rm{erg} \; \rm{cm}^{-2}$] &	& 	3.33$_{-0.05}^{+0.05}$	& 	3.08$_{-0.10}^{+0.10}$	& 	2.50$_{-0.10}^{+0.10}$	& 	2.44$_{-0.08}^{+0.05}$	\\

\hline\\[2pt]
PG-stat / DOF	&				&	440.7 / 354			&	405.5 / 352			&	409.0 / 353			&	390.0/ 352				\\[2pt]
\hline

\hline\end{tabular}
\caption{Best--fit parameters for all the models for the spectrum obtained in the interval P2, [\Tz{3.0}, \Tz{7.6}]. The reference energy is fixed to 1~MeV. The fluence is evaluated for the 10~keV--10~GeV range.}
\label{Tab:lat05gbmt95}
\end{table*}

\begin{figure}[htbp]
  \centering
  \includegraphics[width=\textwidth]{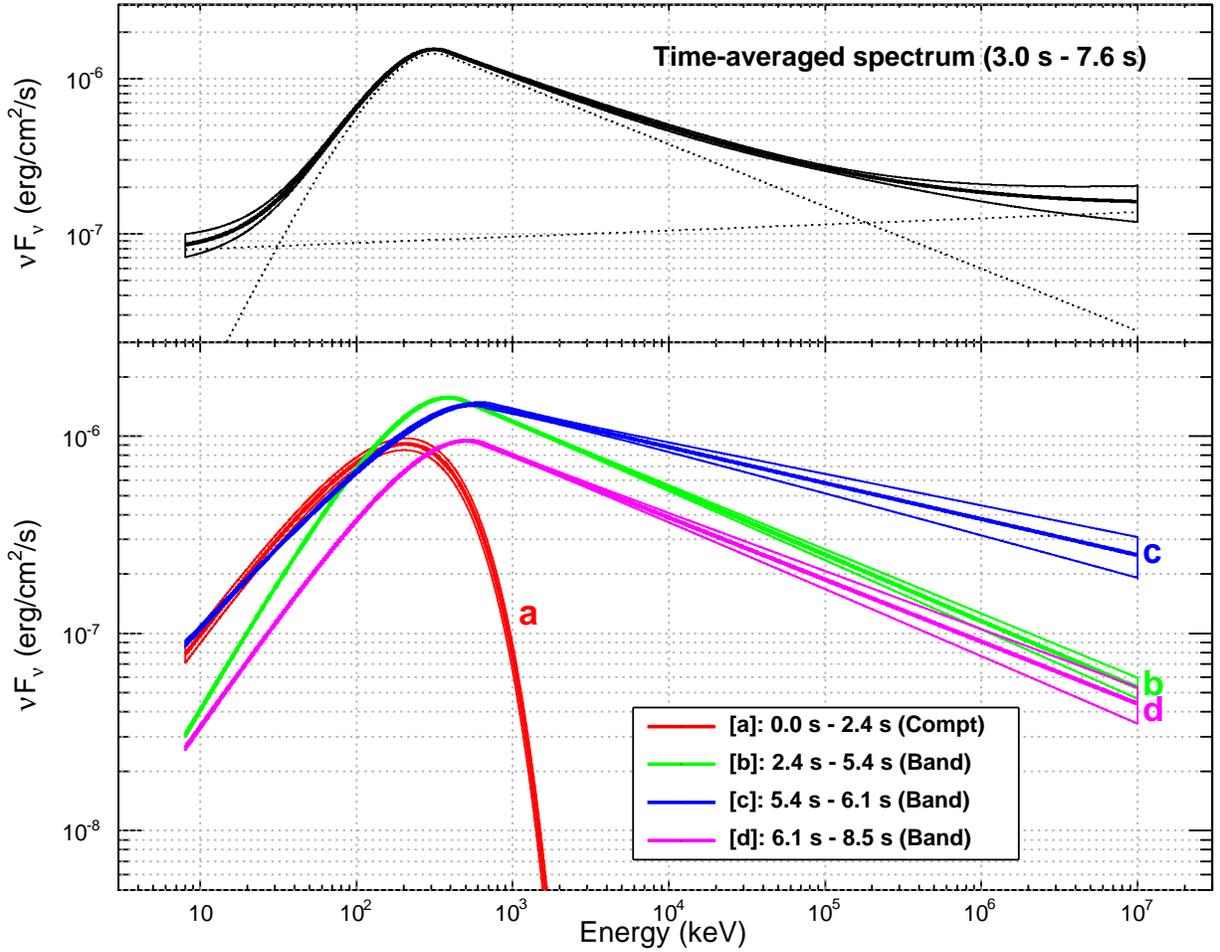}
  \caption{\emph{Top}: The best--fit BAND+PL model for the time--integrated interval P2 plotted as a $\nu F_{\nu}$ spectrum. The two components are plotted separately as the dotted lines, and the sum of the components, representing the overall spectrum, is plotted as the heavy line. The $\pm1~\sigma$ ranges derived from the errors on the fit parameters are also shown. \emph{Bottom}: The $\nu F_{\nu}$ model spectra (and $\pm1~\sigma$ error contours) plotted for each of the time bins considered in the time--resolved spectroscopy.}
  \label{Fig:nuFnu}
\end{figure}

We performed the time--resolved analysis in the 4 intervals $a,b,c,d$ determined by the Bayesian blocks analysis (see Section~\ref{subsection:lightcurves}): a Bayesian blocks algorithm ensures that the flux has no statistically significant variation over each interval, and hence provides a useful binning for studying the spectral evolution of the burst, in particular in the MeV--to--GeV range because the algorithm was run on the LLE light curve. We investigated in particular the presence of the power law component and of a high--energy cutoff. The best--fit spectral parameters for the preferred models are reported in Tab.~\ref{Tab:resolved}, and the best models both for the time--resolved and time--integrated analyses are shown in Fig.~\ref{Fig:nuFnu}. The power law component is not detected above the probability threshold in any of the intervals, likely due to the limited statistics. The results suggest, though, that a below--threshold power--law component may be present starting after the second interval. There is also marginal evidence of a high--energy cutoff in the second interval. Here we summarize the results from the time-resolved spectral studies:
\begin{itemize}
\item [-] {interval $a$ ([0, \Tz{2.44}]):  the only model that was constrained by the data is COMP. }
\item [-] {interval $b$ ([\Tz{2.44}, \Tz{5.44}]): Adding a PL to BAND gave an improvement $\Delta S = 19.5$. Adopting BAND as null hypothesis, the corresponding probability is $P(\Delta S) \simeq 2\times 10^{-5}$, very close to the significance threshold. COMP+COMP marginally improved the fit with respect to BAND+PL, giving $\Delta S=8.1$. Adopting the latter model as null hypothesis, we obtain $P(\Delta S) \simeq 3 \times 10^{-4}$.}
\item  [-]{interval $c$ ([\Tz{5.44}, \Tz{6.06}]): BAND, BAND+PL and COMP+COMP gave very similar values for $S$. Thus, from a statistical point of view,  the more complex models are not favored over BAND. However, it is noteworthy that the best--fit parameters of the complex models gave results which are consistent with what we found for interval $b$. }
\item  [-]{interval $d$ ([\Tz{6.06}, \Tz{8.52}]) : BAND+PL fit the data slightly better than BAND alone, with $\Delta S = 13.7$. Adopting BAND as null hypothesis, we obtain $P(\Delta S) \simeq 3 \times 10^{-4}$. The high--energy cutoff of the COMP+COMP model is not well constrained.}
\end{itemize}

\begin{table*}[htbp]
\centering
\begin{tabular}{lllll}

\hline
Time interval from $T_0$ [s]		&	\emph{a} (0--2.44)				&	\emph{b} (2.44--5.44)                        &	\emph{c} (5.44--6.06)			&                        \emph{d} (6.06--8.52)  \\

\hline
\hline
		
Best model					&	COMP						&		BAND					&	BAND					&	BAND				\\[2pt]
\hline\\[2pt]
$E_0$ [keV]					&	188$_{-17    }^{+22  }$			&	285$_{-26   }^{+30   }$			&	683$_{-180   }^{+270   }$ 		&	446$_{-72   }^{+91   }$		\\[2pt]
$\Gamma_\alpha$					&	 $-0.92_{-0.05 }^{+0.05 }$			&	$-0.64_{-0.05  }^{+0.05  }$		& 	$-1.15_{-0.06 }^{+0.05 }$		&	 $-0.86_{-0.06  }^{+0.06  }$	\\[2pt]
$\Gamma_\beta$						&	$-$							&	$-2.34_{-0.04  }^{+0.04  }$		&	$-2.18_{-0.06  }^{+0.05  }$	&	$-2.31_{-0.05}^{+0.04}$		\\[2pt]
\hline\\[2pt]

Fluence [$10^{-5}\, \rm{erg} \; \rm{cm}^{-2}$]	& 	0.58$_{-0.06}^{+0.05}$		& 	2.05$_{-0.04}^{+0.04}$			& 	0.59$_{-0.03}^{+0.03}$		& 	1.10$_{-0.04}^{+0.04}$		\\[2pt]

\hline\\[2pt]
PG-stat	(DOF)					&								&								&							&							\\

BAND (354)				& 	$-$							&	417.4 						&	 365.3					&	389.1 					\\
COMP (353) 				&	378.6  						&	$-$							&	 $-$						&	$-$						\\
BAND+PL	(352)			& 	$-$							&	397.9 						&	 363.2					&	375.4 					\\
COMP+PL	 (353)			& 	$-$							&	399.7 						&	 365.3 					&	380.2					\\
COMP+COMP (352)			&	$-$							&	389.8 						&	 360.2 					&       377.7 					\\
\hline

\hline\end{tabular}
\caption{Summary of GBM/LAT joint spectral fitting by best model in 4 time intervals. The fluences are evaluated for the range covered by both instruments is 10~keV -- 10~GeV.}
\label{Tab:resolved}
\end{table*}

\subsection{Afterglow modeling}\label{sec:resultsafterglow}

\subsubsection{Multiwavelength afterglow light curves}
\label{sec:mwllc}
\begin{figure*}[htbp]
  \begin{center}
    \includegraphics[width=\textwidth]{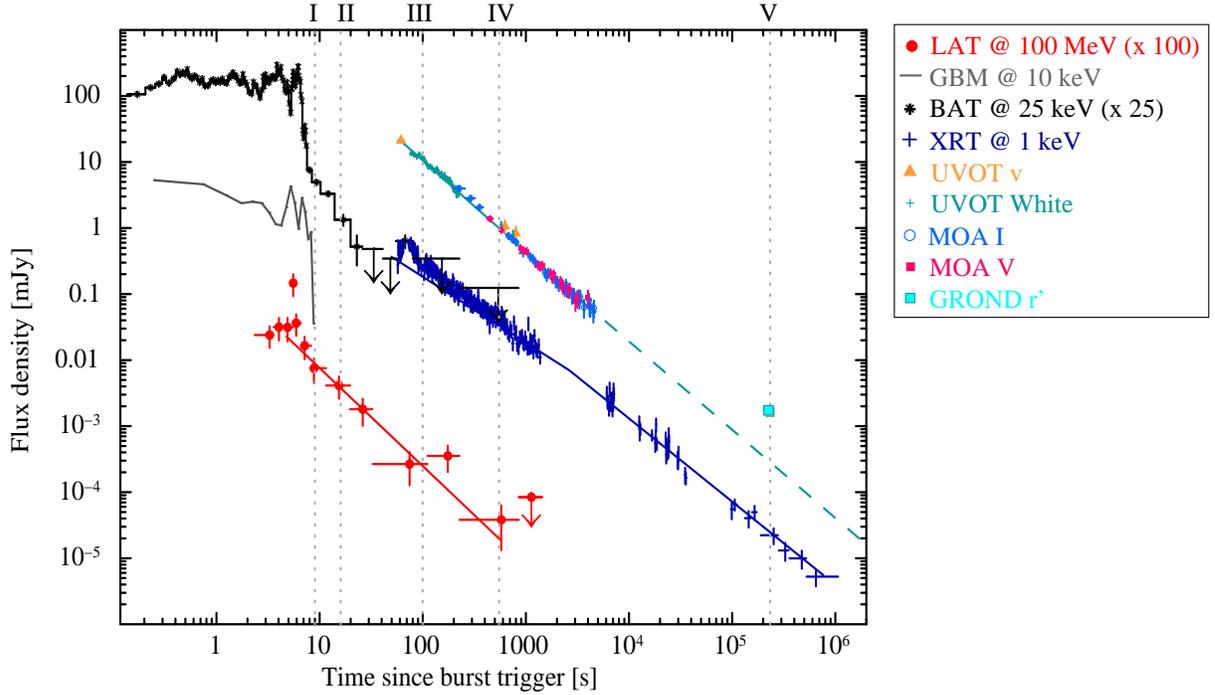}
    \caption{Multiwavelength light curves of \GRB observations: vertical dotted
    lines define the time boundaries of the SEDs (indices I, II, III, IV and V) studied in Section~\ref{sec:sed}.
    The power-law index of the LAT light curve is $\alpha_{\rm LAT} = 1.55\pm 0.20$. The XRT light curve follows a broken power law with $\alpha_{X,1}$=1.10$\pm$0.02, $\alpha_{X,2}$=1.32$\pm$0.03 and a break at $t_{bk}$=4.6$^{+2.6}_{-1.6}$~ks. The UVOT light curve is well fit by a single power law with decay slope $\alpha_{\rm opt}$=1.37$\pm$0.03.}
    \label{Fig:mwllc}
  \end{center}
\end{figure*}

The multiwavelength light curves for the 7 instruments (LAT, GBM, XRT, BAT, UVOT, MOA and GROND) that observed \GRB are shown in Fig.~\ref{Fig:mwllc}, and the corresponding data points are reported in Tab.~\ref{Tab:LATExt} through Tab.~\ref{Tab:grond} in appendix.

The LAT light curve in Fig.~\ref{Fig:mwllc} shows that the peak of the flux density is in the interval [\Tz{5.47},\Tz{5.67}], and that after \Tz{5.67}, the flux decays smoothly during the whole temporally extended emission, following a power law with a fitted index $\alpha=1.55\pm0.20$. The last time interval with a clear detection in the LAT data spans from \Tz{227.0} to \Tz{853.9}.

The BAT light curve shows many consequent adjacent peaks,
from the trigger time to $\sim$\Tz{8.5}.
Faint, spectrally soft emission is detected out to $\sim$\Tz{30}, 
and its temporal decay can be described by a power law with index $\alpha_{\rm BAT}$=2.3$\pm$0.3.
By shifting the reference time to the peak of the observed LAT emission
at t$_{pk}$=\Tz{5.5}, i.e. shifting the reference to the onset of the forward shock, the decay slope changes to $\alpha_{{\rm BAT},t_{pk}}$=1.25$\pm$0.15.
A rebrightening is visible between \Tz{50} and \Tz{80}, 
in coincidence with the first X--ray flare as seen on the multiwavelength light curve (Fig.~\ref{Fig:mwllc}). 

The X--ray light curve shows two X--ray flares, peaking at \Tz{70} and 
\Tz{110} respectively. By excluding the interval of significant variability
(from \Tz{56} to \Tz{150}) from the temporal fit, the light curve can be described by a simple power law
with decay slope $\alpha_{X}=1.189\pm0.007$ ($\chi^2$=250 for 228 d.o.f.).
This model, though acceptable at a statistical level, systematically overestimates
the late time flux densities. A broken power--law yields a significantly better result 
($\chi^2=217$ for 226 d.o.f.) and shows no systematic trend in the residuals.
The best--fit parameters are $\alpha_{X,1}=1.10\pm0.02$, $\alpha_{X,2}=1.32\pm0.03$ and $t_{bk}=4.6^{+2.6}_{-1.6}$~ks. 

For the optical emission, the light curve we obtained using UVOT and MOA data altogether is very well fit  by a single power law with decay slope $\alpha_{\rm opt}=1.37\pm0.03$, as shown in Fig.~\ref{Fig:mwllc}. For the UVOT data, we created a single filter light curve, the \emph{White} light curve, by co-adding normalized version of the light curves for the 7 UVOT filters. The multiwavelength light curve in Fig.~\ref{Fig:mwllc} shows just UVOT \emph{White} and {\em v} data for clarity, as well as MOA data for the {\em I} and {\em V} bands. Data points for individual and co-added exposures are reported in Tab.~\ref{Tab:uvot:uvb} and Tab.~\ref{Tab:uvot:white}.

\subsubsection{Broadband SED}
\label{sec:sed}

We evaluated the SED at 5 epochs: \Tz{9} (I), \Tz{16} (II),
\Tz{100} (III), \Tz{550} (IV), and $2.74$~d (V). The SED times were selected in
order to maximize the data coverage and minimize interpolation. For each epoch and instrument, a spectrum  was extracted using the corresponding time interval, and scaled to the actual count rate at each time of interest. The time intervals were chosen in order to achieve a significant detection in the LAT energy band (epochs I-III), sufficient statistics in the X--ray spectrum (epoch IV) and to just match GROND observations for epoch V. Optical fluxes were interpolated, when necessary, by using the  best--fit model for the light curve. At the GRB redshift of $z=2.83$ the Lyman series absorption is redshifted to the observer frame wavelength range 3500--4700~\AA. This mainly affects the UVOT $b$, $u$, and the GROND $g$' data points, which were therefore excluded from the spectral fits.

Each SED was fit with an absorbed power law or an absorbed broken 
power law, where the spectral slopes $\beta_1$ and $\beta_2$ were tied 
to $\beta_2=\beta_1-0.5$, as predicted by the closure relations for a cooling break~\citep{2004IJMPA..19.2385Z}. The Galactic X--ray absorption and reddening were fixed to the values corresponding to \nh=$1.0 \times$ \e{21}\,\cm{-2} and $E(B-V)$=0.18 \citep{swiftgcnreport}, respectively. 
The intrinsic X--ray absorption was modeled by assuming an absorber with solar
metallicity. To model the host intrinsic extinction we tested each of the
Milky Way (MW), Large Magellanic Cloud (LMC) or Small Magellanic Cloud (SMC) extinction law, as parametrized by \citet{1992ApJ...395..130P}. Limited statistics of our data did not permit any of these laws to be excluded or preferred.

\begin{figure}[htbp]
  \begin{center}
    \includegraphics[angle=-90, width=0.9\textwidth]{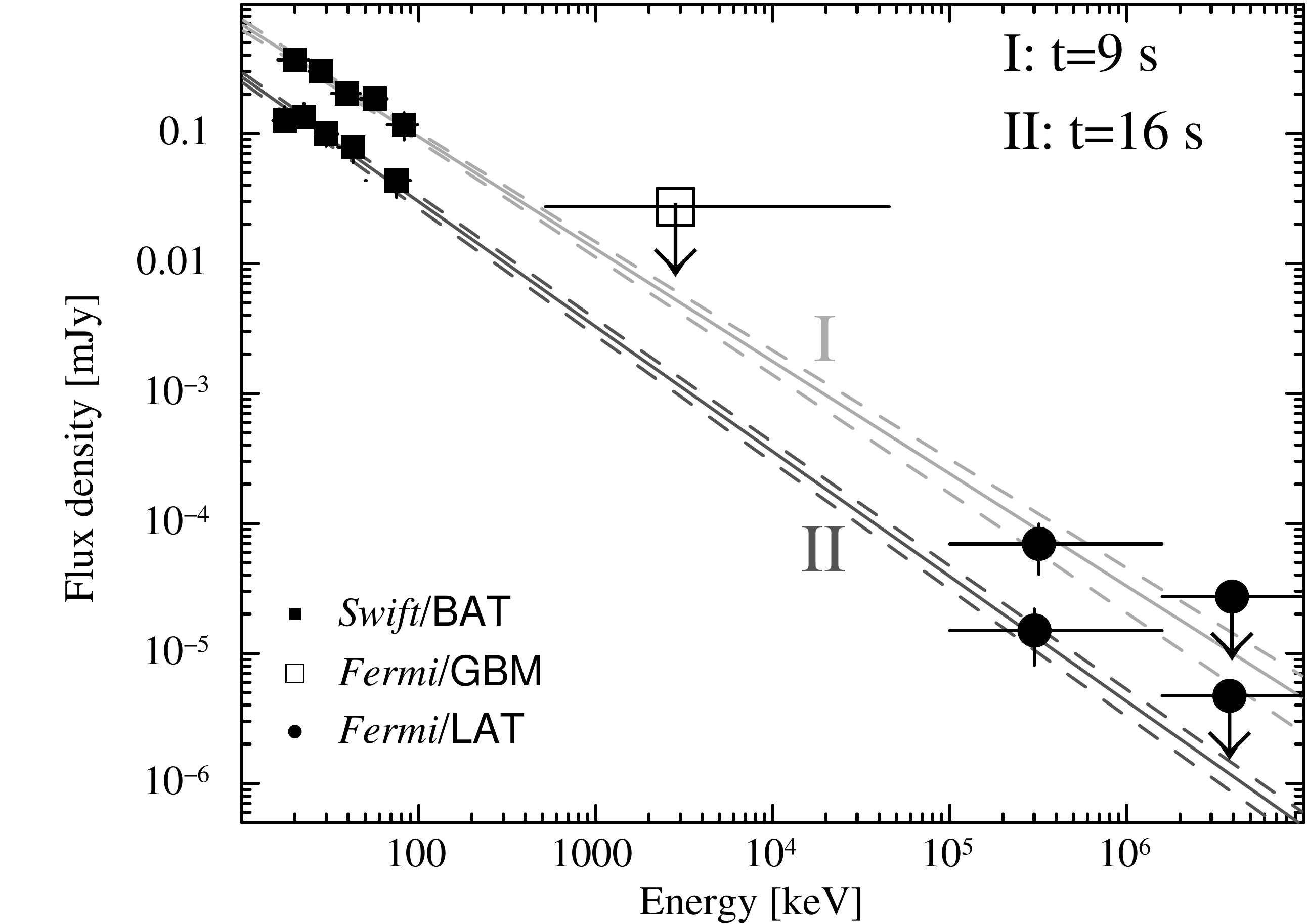} 
    \caption{SEDs at epochs I and II. GBM BGO
    data upper limits at both epochs are very similar, so for clarity only
    the upper limit at epoch I is shown. The SEDs at epochs I and
    II are well fit by a power law with photon indices, $1.87^{+0.07}_{-0.11}$ and $1.95^{+0.07}_{-0.11}$, respectively. The continuous lines represent the best--fit, and the dashed lines the $1~\sigma$ ranges.}
    \label{Fig:seds12}
  \end{center}
\end{figure}

For epochs I and II, the SEDs are well fit by a single power law spectrum over the whole energy range from the keV (BAT data) to the GeV (LAT data), as shown in Fig.~\ref{Fig:seds12}.
 
For epoch III the analysis was complicated by the presence of X--ray flares 
and also limited by the poor optical coverage at early times (c.f.
Fig.~\ref{Fig:mwllc}). In order to estimate the afterglow underlying the
observed flaring activity, we extracted the X--ray spectrum in interval [\Tz{160}, \Tz{227}] and rescaled it by extrapolating the light curve best--fit model. We extracted the spectrum during the X--ray flares from \Tz{56} to \Tz{150}, and subtracted from the estimated afterglow contribution. 

\begin{figure}[htbp]
  \begin{center}
    \includegraphics[angle=-90, width=0.9\textwidth]{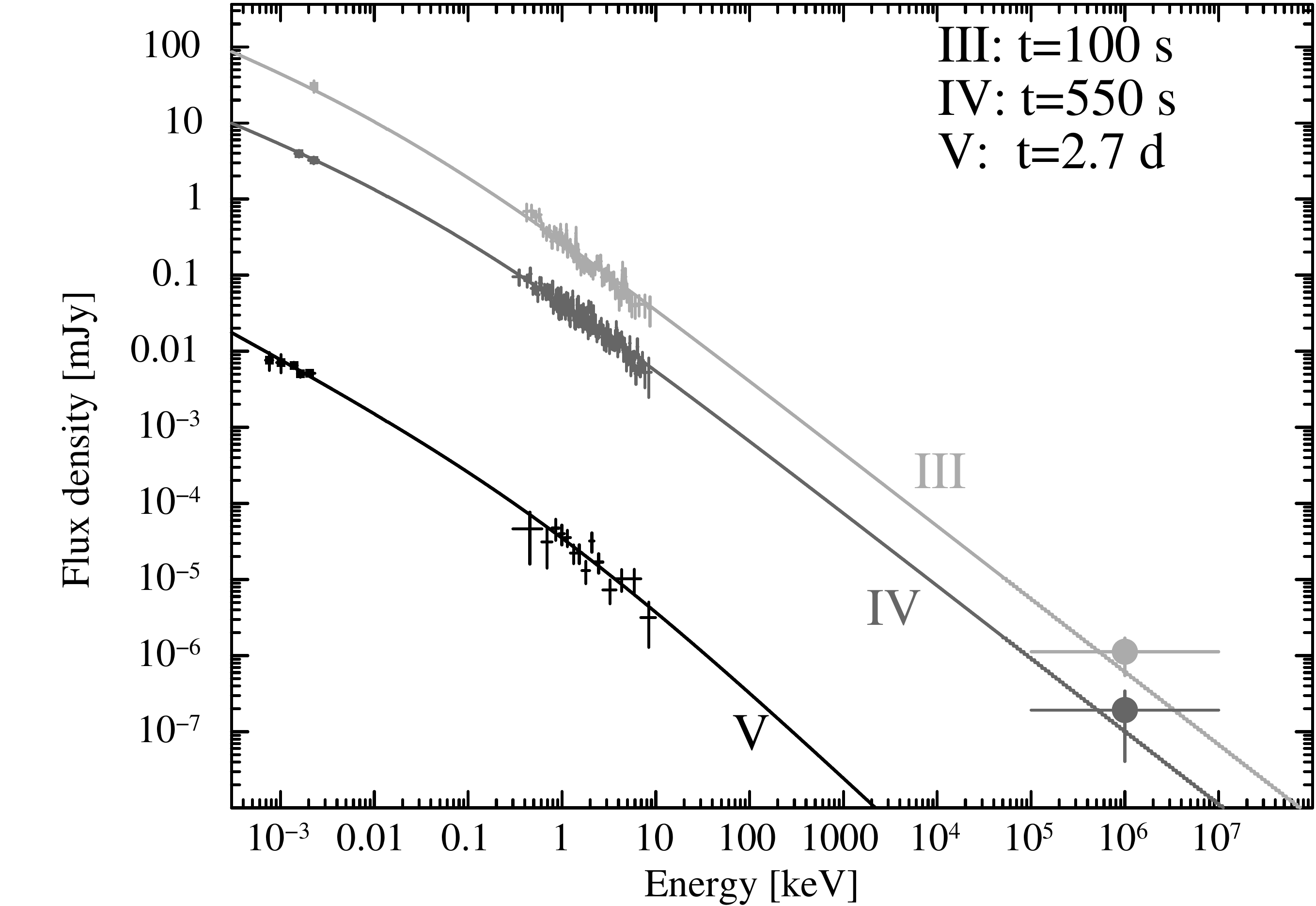} 
    \caption{SEDs for epochs III, IV and V. For epochs III and IV, the spectra are built using UVOT ($\sim3\times10^{-3}$~keV), XRT ($\sim0.5-10$~keV) and LAT ($10^5$--$10^7$~keV) data. For epoch V, we have used the late time GROND observations ($\sim3\times10^{-3}$~keV) and XRT data. The solid lines show the fit of the wind model discussed in the text at each time, for the parameters reported in Tab.~\ref{Tab:SED}.}
    \label{Fig:seds45}
  \end{center}
\end{figure}

We also explored whether the observed SED evolution was consistent
with the expected afterglow behavior in a wind--like density profile~\citep{2002ApJ...568..820G}. 
To this aim, we performed a joint fit by tying some parameters between
the different epochs: we constrained the break energy to increase with time
as $E_{bk}\propto t^{0.5}$, and also held fix the host absorption and extinction as they are not expected to vary. 
We assumed a realistic model for the afterglow spectral shape, 
that is a smoothly broken power law of the form: 
\begin{equation}
F_\nu(E) \propto \left[ (E/E_{bk})^{-s \beta_1} + (E/E_{bk})^{-s \beta_2} \right]^{-1/s}
\end{equation}
where the curvature parameter $s$ was held fixed at $0.8$.
Epoch III was excluded from the fit procedure, but compared to the resulting best--fit model. SEDs and best--fits using the wind model for epochs III, IV and V are shown in Fig.~\ref{Fig:seds45}, and the best--fit results reported in Tab.~\ref{Tab:SED} for all five epochs.

\begin{table*}[htbp] 
\centering 
\begin{tabular}{lccccc} 
\hline\hline 
Model/Epoch   & I (9~s)      &      II (16~s)   &  III (100~s)   & IV (550~s) & V (2.7~d)  \\ \hline\hline 
power law        &          &             &            &      &      \\
 Index            & $-1.87_{-0.07}^{+0.11}$ & $-1.95_{-0.07}^{+0.11}$ & --- & --- & --- \\ \hline
Broken power law &          &             &            &      &      \\
$\beta_{opt}$     & --- & --- & --- & $-0.45^{+0.07}_{-0.09}$ & $-0.66^{+0.03}_{-0.03}$ \\
$\beta_{X}$       & --- & --- & --- & $-0.95^{+0.07}_{-0.09}$ & $-1.16^{+0.03}_{-0.03}$ \\
$E_{break}$ (keV) & --- & --- & --- & $0.04^{+0.03}_{-0.01}$ & $\sim0.8$                \\
s (curvature)     & --- & --- & $0.8$ & $0.8$  & $0.8$             \\
 $\chi^2$/d.o.f   & --- & --- & --- & \multicolumn{2}{c}{$654/731$}                     \\ \hline\hline 
 \end{tabular} 
\caption{Results of broadband fits of the SEDs.
For epochs I and II the simple power law is the best model.
Epochs IV and V have been fitted together with a smoothly broken power law,
with a curvature $s=0.8$, $E_{break}\sim t^{0.5}$ and $\beta_{opt}$=$\beta_{X}$+0.5. Epoch III was not part of the fitting procedure but we check that the data were compatible with the best--fit model obtained for epochs IV and V.}
\label{Tab:SED} 
\end{table*} 



\section{Discussion and Interpretation}
\label{sec:disc}
\subsection{Prompt emission}
\label{sec:disc:prompt}
GRB~110731A was bright in the LAT data with its most prominent peak in interval \emph{c}. The onset of the LAT emission is delayed by 2.5 seconds with respect to the GBM $T_0$, a time delay comparable to the few--seconds delay observed in most long--duration LAT GRBs. The time--resolved spectral analysis suggests that an additional, hard emission component becomes increasingly prominent with time.  Indeed, in \emph{b}, the BAND+PL model fits the spectrum better than a simple BAND, with a significance only slightly below our threshold.  While an additional component cannot be statistically resolved in \emph{c}, an additional power law component improves the BAND--only fit in \emph{d} as well, although with a rather low significance. By adopting the BAND+PL model for intervals \emph{b} and \emph{d} we can measure the ratio of the fluences of the power--law and BAND components, both measured in the 10 keV -- 10 GeV energy range. The ratio is 20\% in interval \emph{b} and 35\% in interval \emph{d}, respectively. 
This trend confirms an emerging picture, corroborated by spectral analysis of other bright LAT bursts with evidence of a power law component in addition to the BAND component (namely GRB~090510, \citealp{2010ApJ...716.1178A}; GRB~090902B, \citealp{2009ApJ...706L.138A} and GRB~090926A, \citealp{2011ApJ...729..114A}), for which the flux of the additional power law component grows with time, and most likely dominates in the temporally extended LAT emission beyond the prompt emission phase.  

The origin of this additional spectral component is not yet fully understood.  An early afterglow model~\citep{2009MNRAS.400L..75K,MNR:MNR17274,2010MNRAS.403..926G, 2010ApJ...709L.146D, 0004-637X-720-2-1008, 2010ApJ...724L.109R}, that is consistent with the above scenario, produces the power law component from the forward shock of the GRB blast wave that propagates into the external medium surrounding the GRB~\citep{1997ApJ...476..232M,1998ApJ...497L..17S}. The delayed onset of the LAT emission is explained as the time required for the forward shock emission to become detectable in this scenario.
In the context of the internal shock model~\citep{1994ApJ...430L..93R}, the additional spectral component can arise due to Compton scattering of soft target photons by relativistic electrons~\citep{1994MNRAS.269L..41M,2009ApJ...698L..98W,2009A&A...498..677B,MNR:MNR18807}. The $2.44$~s delay for the LAT--detected emission in these scenarios would indicate GeV emission to be variable on a similar time scale and arising due to Compton emission from late internal shocks.  Finally, hadronic emission, either proton/ion synchrotron radiation or photo--pion--induced cascade radiation, can produce an additional spectral component~\citep{2009ApJ...705L.191A,2010OAJ.....3..150R,2009ApJ...698L..98W}.

The delayed onset of the high energy emission in the hadronic models is interpreted as due to the time required for proton/ion acceleration and cooling, as well as to the time required to form cascades. In particular, to generate the $2.44$~s delay in LAT the jet magnetic field needs to be $B^\prime \approx 10^5 (\Gamma_0/500)^{-1/3} (t_{\rm onset}/2.44~{\rm s})^{-2/3} (E_\gamma/100~{\rm MeV})^{-1/3}$~G, with a jet bulk Lorentz factor $\Gamma_0$, in case of proton--synchrotron model and the corresponding isotropic--equivalent jet luminosity is $L_{\rm jet} \gtrsim 10^{57} (\Gamma_0/500)^{16/3} (t_{\rm onset}/2.44~{\rm s})^{2/3} (E_\gamma/100~{\rm MeV})^{-2/3}$~erg~s$^{-1}$~\citep{2010OAJ.....3..150R,2009ApJ...698L..98W}. While the hadronic models require a larger total energy than the leptonic models, the energy budget can be brought down to acceptable levels for a bulk Lorentz factor $\lesssim 500$~\citep{2010ApJ...716.1178A, 2011ApJ...729..114A}. In addition, the flat spectrum (index $\sim -2$) for the extra component is also favorable to the energy budget, and the observed low--energy (10 keV) excess (c.f. Fig.~\ref{Fig:nuFnu}, top) could be additional evidence for a hadronic model.  These two properties are very similar to the features already reported for the LAT burst GRB~090902B~\citep{2009ApJ...706L.138A, 2010ApJ...725L.121A}.

At a redshift $z=2.83$, the peak isotropic--equivalent luminosity (10~keV--10~GeV) of \GRB, as measured in interval \emph{c}, is $L_{\rm iso,pk} = 6.3^{+0.3}_{-0.3} \times 10^{53}$~erg~s$^{-1}$. The corresponding peak of the energy spectrum is at $E_{\rm pk} = (2+\Gamma_\alpha)E_0 = 683^{+270}_{-180}$~keV, from the BAND fit.  The total isotropic--equivalent energy (10~keV--10~GeV) over the full prompt phase (P1) is $E_{\rm iso} = 6.0^{+0.1}_{-0.2} \times 10^{53}$~erg, assuming a COMP+PL model. The BAND+PL fit to the same spectrum results in a somewhat larger $E_{\rm iso} = 7.6^{+0.2}_{-0.2} \times 10^{53}$~erg with a BAND $E_{\rm pk} = (2+\Gamma_\alpha)E_0 \approx 300$~keV.  These values are broadly consistent with the empirical $L_{\rm iso,pk}$--$E_{\rm pk}$~\citep{2004ApJ...609..935Y} and  $E_{\rm iso}$--$E_{\rm pk}$~\citep{2009A&A...508..173A} relations for long GRBs. Another empirical relation between the luminosity and the spectral lags that we measured to be $-41\pm 28$~ms between two \Swift BAT energy bands, is also satisfied~\citep{2000ApJ...534..248N,MNR:MNR19723}.

Constraining the bulk Lorentz factor, $\Gamma_0$, of the jet is a major challenge in GRB science.  A constraint from the $\gamma\gamma$ pair production opacity argument can be used to derive a lower limit on $\Gamma_0$ if the spectrum does not have a cutoff and  extends to the highest observed energy~\citep{1991ApJ...373..277K,1993A&AS...97...59F,1997ApJ...491..663B,2001ApJ...555..540L,2004ApJ...613.1072R}. For \GRB, a 2.0~GeV photon was detected at \Tz{8.27}.  The spectrum in interval \emph{d} which included this photon, is preferably fit with a BAND+PL.  We derive a lower limit, $\Gamma_{\rm min} \sim 600$, from $\tau_{\gamma\gamma} < 1$ using the above information and the median value of the FWHM of the pulses within the $T_{90,\rm{GBM}}$, $\Delta t = 0.35\pm 0.02$~s, as the variability time in the single--zone emission approximation with a homogeneous distribution of photons~\citep[for a description of the method, see][]{2010ApJ...716.1178A}.  The corresponding emission radius is $R \gtrsim \Gamma_{\rm min}^2 c\Delta t (1+z)^{-1} \sim 10^{15}$~cm. In interval P2 we detect a high--energy cutoff with a significance only slightly below our threshold. This provides an opportunity to estimate $\Gamma_0$ as in GRB~090926A, by assuming that the cutoff is due to $\gamma\gamma$ pair production~\citep{2011ApJ...729..114A}.  We derive $\Gamma_0 = 530\pm 10$, using the COMP+COMP model with a $390^{+220}_{-120}$~MeV folding energy for the high--energy component, and with $\Delta t = 0.43\pm 0.03$~s, the median value of the FWHM of the pulses in P2.  The maximum photon energy in this time interval is 0.9~GeV. Observe that while regions somewhat smaller than those inferred from these choices of $\Delta t$ can be possible, since the pair opacity is a strong function of $\Gamma_0$, the values of the bulk Lorentz factor inferred here are only weakly dependent on the choice of $\Delta t$. Note that effects within the emission region such as radiation transport~\citep{2010ApJ...716.1178A} or a time--dependent increase of $\tau_{\gamma\gamma}$ and geometrical effects~\citep{2008ApJ...677...92G, 2012MNRAS.421..525H} can reduce the value of $\Gamma_{\rm min}$ or $\Gamma_0$ by a factor of up to 3, even for a single--zone model.  Indeed the $\Gamma_{\rm min}$ for the interval {\it d} reduces to a lower limit of $\sim 300$ for the parametrization by~\cite{2012MNRAS.421..525H}. Limits on the bulk Lorentz factor can be further relaxed for two--zone emission models, where GeV photons are emitted from a larger radius than the MeV photons~\citep{2011ApJ...726L...2Z,2012MNRAS.421..525H}. Such reductions in $\Gamma_0$ clearly can prove important for the viability of hadronic models, as discussed above.

\subsection{LAT temporally extended emission and multiwavelength afterglow}\label{sec:extemmwag}

Thanks to the wealth of simultaneous multiwavelength data, we can test various afterglow models and extract the preferred model parameters. In particular, these observations can help constrain whether the temporally extended LAT emission originates in the forward shock that produces the late--time afterglow emission.

We discuss epoch IV first (c.f. Fig.~\ref{Fig:mwllc}) since the data are most constraining in this time interval.  In Section~\ref{sec:swift} we observed that the optical flux ($\alpha_{\rm opt} = 1.37\pm 0.03$) decays faster than the X--ray flux ($\alpha_{X,1} = 1.10\pm 0.02$ or $\alpha_X = 1.189\pm 0.007$), and in Section~\ref{sec:sed}, we reported that the broadband SED is well fit by a smoothly broken power law at epoch IV. Both features favor a wind rather than ISM afterglow model, with a slow--cooling spectrum~\citep{2000ApJ...536..195C,2000ApJ...543...66P}.  This results from an examination of the relations between the decay and spectral indices that describe flux evolution, $F_\nu \propto t^{-\alpha} \nu^{-\beta}$, at different energies of the afterglow synchrotron spectra~\citep{2002ApJ...568..820G,2006ApJ...642..354Z}.  The break in the SED at epoch IV can be readily interpreted as a cooling break ($\nu_c$) rather than a break due to the typical synchrotron frequency of the minimum energy electrons ($\nu_m$), in both the slow-- and fast--cooling spectra; the respective synchrotron indices below such
breaks are $\beta =2/3$ (slow--cooling) or $\beta = 3/2$ (fast--cooling).  This is because the fast decay of the observed optical flux is contrary to (i) an increasing (ISM) or invariant (wind) flux behavior expected below $\nu_m$ in the slow--cooling spectrum; and (ii) a much slower decay behavior, $F_\nu \propto t^{-1/4}$ (ISM or wind), expected below $\nu_m$ in the fast--cooling spectrum. Further investigation shows that the fast--cooling spectrum is disfavored at epoch IV, since (i) $\alpha_{X,1}$ or $\alpha_X$ and $\beta_{X,1}$ ($= 0.87\pm 0.04$, derived as $\Gamma_X -1$ before the temporal break) for the X--ray flux are both incompatible with the $F_\nu \propto t^{-1/4}\nu^{-1/2}$ behavior expected for $\nu_c < \nu_X < \nu_m$ (ISM or wind); and (ii) $\alpha_{\rm opt}$ is incompatible with either $F_\nu \propto t^{1/6}$ (ISM) or $F_\nu \propto t^{-2/3}$ (wind) behavior for $\nu_{\rm opt} < \nu_c$.  The spectral index $\beta_{X,\rm SED} = 0.95^{+0.09}_{-0.07}$ for $\nu_c < \nu_X$, fitting the broadband SED at epoch IV, is compatible, within 1--2~$\sigma$, with $\beta_{X,1}$ and with the expected $\beta = (2/3)\alpha +1/3$ behavior in the slow--cooling spectrum for $\alpha_{X,1}$ or $\alpha_{X}$.  While the above is common behavior for both the ISM and wind models, the observed $\alpha_{\rm opt}$ favors a $\beta = (2/3)\alpha -1/3$ relation (wind) over a $\beta = (2/3)\alpha$ relation (ISM) when compared with the broadband SED fit result $\beta_{\rm opt,SED} = \beta_{X,\rm SED} -0.5$.  The expected optical spectral index is within 2~$\sigma$ of $\beta_{\rm opt,SED}$ in the wind model and deviates by more than 5~$\sigma$ from $\beta_{\rm opt,SED}$ in the ISM model.  Thus our analysis favors a slow--cooling spectrum and wind environment for broadband data at epoch IV.

The broadband SED at 2.74~d, epoch V, has far less predictive power than the SED at epoch IV, with a single power law being statistically favored over a broken power law.  Nevertheless, the joint fit to both SEDs using a smoothly broken power law, with tied indices but allowing the break frequency to increase as $\propto t^{1/2}$ between the SEDs as expected in the wind model, resulted in a good fit.  This result further supports the wind model.  A steeper decrease in the X--ray flux ($\alpha_{X,2} = 1.32\pm 0.03$) after $4.6^{+2.6}_{-1.6}$~ks can be explained as the break frequency $\nu_c$ approaching the X--ray band with time and reaching $\sim 1$~keV at 2.74~d.  An optical rebrightening is clearly visible in GROND data at epoch V without an accompanying rebrightening in X--ray, as observed in GRB~081029~\citep{2011A&A...531A..39N}. The optical rebrightening could possibly originate from a two--component jet structure as modeled in GRB~030329~\citep{2003Natur.426..154B} and GRB~080319B~\citep{2008Natur.455..183R}. However, the lack of data during the rise of the optical flux and sparse late--time data prohibit further study of the optical rebrightening in GRB~110731A. 

The LAT temporally extended emission that is contemporaneous with the BAT and GBM detected emission at very early times, epochs I and II, can be fit with single power--law models to produce a broadband SED.  The resulting  photon indices, $1.87^{+0.07}_{-0.11}$ and $1.95^{+0.07}_{-0.11}$ respectively, are compatible with the measured X--ray photon index $\Gamma_X$ at later times.  In the context of the wind afterglow model, these two very early SEDs can be interpreted as being due to emission above $\nu_m$ in a fast--cooling spectrum ($\nu_c < \nu_m$) which subsequently turns into a slow--cooling spectrum ($\nu_m < \nu_c$) as $\nu_c$ increases with time and moves past $\nu_m$, giving rise to the broken power law SED at epoch IV.  However, while the LAT flux decay index $\alpha_{\rm LAT} = 1.55\pm 0.20$ is marginally consistent within 2~$\sigma$ with the later X--ray flux decay index $\alpha_{X,1}$ or $\alpha_X$, the BAT flux decays ($\alpha_{\rm BAT} = 2.3\pm 0.3$) at a much faster rate.  A steeper decay of the BAT flux can be due to an emission component, such as high--latitude ($\theta > \Gamma_0^{-1}$) emission from the fireball~\citep{2000ApJ...541L..51K,2004ApJ...614..284D}, in addition to the underlying afterglow emission.  Indeed, fitting the BAT flux as $F_\nu \propto (t-t_{\rm pk})^{-\alpha}$, where $t_{\rm pk} = 5.5$~s was chosen to coincide with the brightest peak in the LAT light curve, results in a flux decay index ($\alpha_{{\rm BAT},t_{\rm pk}} = 1.25\pm 0.15$) which is more compatible with $\alpha_{\rm LAT}$ and $\alpha_{X,1}$ or $\alpha_X$, yet captures the steep decay of the BAT flux at very early time.  Part of the LAT emission at $\sim$\Tz{5.5} likely originates from such high--latitude emission and/or internal shocks.  Indeed an extrapolation of the LAT afterglow flux at this time fails to reproduce all of the observed emission, as also noted for GRB~090510~\citep{2011ApJ...733...22H}.

Epoch III is rather complex with 2 weak X--ray flares.  The most likely origin of the flares and a rebrightening of the emission detected by the BAT in coincidence with the first flare is late--time activity of the central engine~\citep{2006ApJ...642..354Z}, as also implied by the noticeable variation of the spectrum from the underlying afterglow emission.  The change in the afterglow synchrotron spectrum from the fast-- to slow--cooling regime is expected to take place in this interval or earlier, since the SED in the next interval is best fitted by a broken power law as mentioned above.

Following the above discussion on the broadband SED of GRB~110731A for multiple epochs and flux decay behavior, two main features can be highlighted:
\begin{itemize}
\item LAT temporally extended emission, as early as \Tz{8.3} as seen in the first broadband SED, is compatible with the afterglow synchrotron emission in other bands. 
\item A wind afterglow model is favored over an ISM model, from the behavior of the SEDs.
\end{itemize}
In the wind afterglow scenario, however, the onset of the afterglow needs to take place quite rapidly.  There are several hints that this is the case: (i) the brightest peak in the LAT light curve dominating the $\nu F_\nu$ flux at all frequencies takes place as early as \Tz{5.5}; (ii) the absence of a brighter flux in the BAT or GBM at a later time; and (iii) the LAT flux decays smoothly after \Tz{5.5}. While the last point is quite obvious, the first two hints result from the fact that the bolometric flux of the forward--shock emission peaks at the deceleration time of the GRB fireball~\citep{1976PhFl...19.1130B,1997ApJ...489L..37S,2010MNRAS.403..926G}.  In the case of an adiabatic fireball in the wind environment, the deceleration time is $t_{\rm dec} = (1+z)E_{k}/(16\pi m_pc^3 A \Gamma_0^4)$, where $E_k$ is the isotropic--equivalent kinetic energy of the fireball during the decelerating phase and $A = 3.02\times 10^{35} A_\star$~cm$^{-1}$ is the wind parameter for a $10^{-5} M_\odot$~yr$^{-1}$ mass--loss rate in the wind of velocity $10^3$~km~s$^{-1}$, with $A_\star =1$ as an scaling parameter~\citep{2000ApJ...536..195C}.  Note that formally the afterglow onset requires that $t_{\rm dec} \gtrsim T_{\rm GRB}$, where $ T_{\rm GRB}$ is the duration of the prompt phase, for the ``thin--shell'' formula used here.  Nevertheless, most of the emission detected by the GBM (50--300~keV) comes within the $T_{95,\rm{GBM}}$ ($7.6\pm0.3$~s), which is similar to the deceleration time scale needed to explain the earliest broadband SEDs.  Moreover, a fraction of the emission arriving later in the formal prompt phase may originate from the very early afterglow.  

\subsection{Afterglow parameters in the wind model}
\label{sec:afterglowparam}
Adopting the wind afterglow model and the broadband SED fit parameters of epoch IV, we derive the afterglow model parameters for GRB~110731A.  The optical and X--ray flux decay indices constrain the electron acceleration index to be $p = (4/3)\alpha_{\rm opt} +1/3 = 2.16\pm 0.04$ and $p = (4/3)\alpha_{X,1} +2/3 = 2.13\pm 0.03$, respectively, from the $F_\nu \propto t^{-\alpha} \nu^{-\beta}$ relations. For the sake of simplification, we assume $p=2.2$. We then use the parametrization by~\citet{2002ApJ...568..820G} for the break frequencies ($\nu_m$, $\nu_c$), the fluxes at the break frequencies ($F_{\nu_m}$, $F_{\nu_c}$), and the transition time ($t_0$) from the fast-- to slow--cooling spectrum as
\ba
h\nu_m &=& 3.5\times 10^4~\eps_e^2\eps_B^{1/2} E_{55}^{1/2}t_1^{-3/2} ~{\rm keV},
\nonumber \\
h\nu_c &=& 2.8\times 10^{-8}~\eps_B^{-3/2} A_\star^{-2} E_{55}^{1/2} t_1^{1/2} ~{\rm keV},
\nonumber \\
F_{\nu_m} &=& 7.1\times 10^4~\eps_B^{1/2} A_\star E_{55}^{1/2} t_1^{-1/2}  ~{\rm mJy},
\nonumber \\
F_{\nu_c} &=& 1.2\times 10^{12}~\eps_e^{1.2} \eps_B^{1.7} A_\star^{2.2} E_{55}^{1/2} t_1^{-1.7} ~{\rm mJy},
\nonumber \\
t_0 &=& 1.1\times 10^7~\eps_e \eps_B A_\star ~{\rm s}.
\ea
Here $\epsilon_e$ and $\epsilon_B$ are the usual micro--physical parameters~\citep{1998ApJ...497L..17S,2002ApJ...568..820G}, $E_{55} = E_k/10^{55}$~erg, and $t_1 = t/10$~s.  

Using the reference time $t_c \sim 550$~s for the cooling break in the SED, $h\nu_c \sim 0.04$~keV and $F_{\nu_c} \sim 0.5$~mJy, we obtain acceptable afterglow parameters for $t_{\rm dec} \sim 5.9$~s, the middle of the time interval with the brightest LAT peak, as: 
%
\begin{eqnarray}
E_k = 3.3\times 10^{54} \left( \frac{A_\star}{0.05} \right) \left( \frac{t_{\rm dec}}{5.9~{\rm s}} \right) \left( \frac{\Gamma_0}{500} \right)^4 ~{\rm erg},
\nonumber \\
\epsilon_e = 3.2\times 10^{-3} \left( \frac{A_\star}{0.05} \right)^{-5/6}  \left( \frac{t_{\rm dec}}{5.9~{\rm s}} \right)^{-8/9}  \left( \frac{\Gamma_0}{500} \right)^{-32/9} \left( \frac{t_{c}}{550~{\rm s}} \right)^{17/18} \left( \frac{h\nu_c}{0.04~{\rm keV}} \right)^{17/18} \left( \frac{F_{\nu_c}}{0.5~{\rm mJy}} \right)^{5/6},
\nonumber \\
\epsilon_B = 1.1\times 10^{-2} \left( \frac{A_\star}{0.05} \right)^{-1}  \left( \frac{t_{\rm dec}}{5.9~{\rm s}} \right)^{1/3}  \left( \frac{\Gamma_0}{500} \right)^{4/3} \left( \frac{t_{c}}{550~{\rm s}} \right)^{1/3} \left( \frac{h\nu_c}{0.04~{\rm keV}} \right)^{-2/3},
\nonumber \\
t_0 = 19.5 ~\left( \frac{A_\star}{0.05} \right)^{-5/6}  \left( \frac{t_{\rm dec}}{5.9~{\rm s}} \right)^{-5/9}  \left( \frac{\Gamma_0}{500} \right)^{-20/9} \left( \frac{t_{c}}{550~{\rm s}} \right)^{23/18} \left( \frac{h\nu_c}{0.04~{\rm keV}} \right)^{5/18} \left( \frac{F_{\nu_c}}{0.5~{\rm mJy}} \right)^{5/6} ~{\rm s}.
\label{fits}
\end{eqnarray}
%
Thus the fast-- to slow--cooling transition takes place towards the end of epoch II, in which a single power law is a better fit to the broadband SED.  The coasting bulk Lorentz factor $\Gamma_0 \sim 500$ is compatible with the value/lower limit obtained from the $\gamma\gamma$ opacity argument for the internal shocks. The kinetic energy is also a factor $\sim 5$ larger than the isotropic--equivalent $\gamma$--ray energy $E_{\rm iso} = (6.8\pm 0.1)\times 10^{53}$~erg, which is compatible with the typically assumed ratio between the two energies.  A deceleration time later than \Tz{5.9} but earlier than $\sim$\Tz{55} can also be accommodated with acceptable parameter values in this scenario, without large deviations from the $\Gamma_0 \sim 500$ and $A_\star \sim 0.05$ values used in Eq.~(\ref{fits}).  However, the fact that the broadband SED at epochs I and II  can be fitted with power laws, strongly argues in favor of an afterglow onset time before $8.3$~s.  The variations of the parameter values derived in Eq.~(\ref{fits}) are minimal in this case. With the parameters $E_k$ and $A_\star$ fixed, a change of $t_{dec}$ from \Tz{5.9} to \Tz{8.3} corresponds to $\Gamma_0 \sim 545$. 

The maximum energy of the synchrotron photons, which is independent of the micro--physical parameters~\citep[see, e.g.,][]{2010OAJ.....3..150R}, can be written as:
\ba
h\nu_{\rm max} &\approx & 0.24\,\phi^{-1} (1+z)^{-1} \Gamma~{\rm GeV} 
\nonumber \\
&\sim & 15\,\phi^{-1} t_2^{-1/4} \left( \frac{A_\star}{0.05} \right)^{-1/4} 
\left( \frac{E_k}{10^{54.5}\,{\rm erg}} \right)^{1/4}  ~{\rm GeV}.
\end{eqnarray}
Here $\phi^{-1} \lesssim 1$ is the acceleration efficiency for electrons and $t_2 = t/100$~s.
This equation applies to scenarios where lepton acceleration is gyroresonant and is radiation--reaction limited by synchrotron cooling. In such cases, the maximum electron energy in the comoving jet frame is of the order of $m_ec^2$ divided by the fine structure constant, a result well-known in the context of active galaxies~\citep{1983MNRAS.205..593G} and the Crab nebula~\citep{1996ApJ...457..253D}. Thus detection of a 3.4~GeV photon at $\sim$\Tz{436} from \GRB (see end of Section~\ref{subsection:lightcurves}) is consistent with the maximum synchrotron photon energy for the afterglow parameters derived in Eq.~(\ref{fits}).

\subsection{Comparisons with other GRBs}
The sample of LAT--detected GRBs is still relatively small, with a detection rate of $\sim 8$/yr~\citep{Piron12}.  Here we briefly comment on the properties of \GRB in comparison with other LAT--detected GRBs that were bright at GeV energies, and with GBM--detected GRBs in general. More detailed comparisons will be presented in a catalog of LAT GRBs that is in preparation. We also comment on the X--ray and optical afterglow of \GRB in comparison with other GRBs.

The total (10~keV--10~GeV) isotropic--equivalent energy $E_{\gamma,\rm iso}$ of \GRB falls in the middle of the distribution for LAT bursts with known redshifts; only GRB~100414A, GRB~091003, GRB~090328 and the short burst GRB~090510 have lower $E_{\gamma,\rm iso}$. In the 1 keV--10 MeV range, $E_{\gamma,\rm iso}$  for \GRB is lower than for a number of GBM--only GRBs in the same redshift range.  The fact that \GRB was very close to the LAT boresight played a crucial role for its detection at $\gtrsim 100$~MeV.  An additional power law component as in \GRB has been detected in all bright LAT--detected GRBs and a cutoff in the power law for the $\gtrsim 100$ MeV energy range has been reported in GRB~090926A~\citep{2011ApJ...729..114A} in addition to the present case.  The value of the jet bulk Lorentz factor and its lower limits has been calculated, using $\gamma\gamma$--opacity argument for simple single--zone model, in the range  of $\gtrsim 200$ to $\gtrsim 1000$ for bright LAT GRBs~\citep{2009Sci...323.1688A,Greiner2009A&A,2009ApJ...706L.138A,2010ApJ...716.1178A,2011ApJ...729..114A,Cenko2011ApJ}.  The $\Gamma_0 \sim 500$ value that we derived for \GRB using the same argument is in the middle of this range.  Note that uncertainties in underlying emission modeling may scale down these values by a factor 2--3~(Sec.~\ref{sec:disc:prompt}). Although the current sample is rather small, a cutoff in the BAND or PL spectrum could be one reason for the LAT non--detection of a number of bright GBM bursts within the LAT field of view~\citep{2012ApJ...754..121T}.



Temporally extended high--energy emission detected in \GRB is quite common among LAT--detected GRBs, showing a smooth power law decay of the flux.  The flux decay index, however, is not the same among bursts but varies between $\sim 1$ and $\sim 1.5$~\citep{2010MNRAS.403..926G}.  The common behavior of the temporally extended LAT flux is compatible with afterglow emission, but a direct model comparison with contemporaneous multiwavelength data sets is possible only for the short burst GRB~090510 \citep{2010ApJ...709L.146D,2010ApJ...716.1178A} and for the current long burst \GRB.  

The X--ray afterglow of \GRB does not display the ``canonical'' steep--, flat--, normal--decay behavior~\citep{2006ApJ...642..389N} observed in $\sim$40\% of all long GRBs detected by \Swift. This trend for \GRB is also observed in other LAT bursts with early--time \Swift\ XRT observations such as GRB~100728A~\citep{2011ApJ...734L..27A} and GRB~110625A~\citep{2012ApJ...754..117T}, and may suggest that the afterglows of LAT bursts tend to be dominated by the bright forward--shock emission rather than prolonged episodes of energy injection. 

We have studied the optical afterglow of \GRB using the method of \citet{Kann2006ApJ} and compared it with other well--measured afterglows~\citep{Kann2006ApJ,Kann2010ApJ,Kann2011ApJ}. As the data quality in the optical is not sufficient to allow a fit with all parameters free, we fix the underlying spectral slope to the value derived from the broadband X--ray--to--optical fit, $\beta=0.66$. A free fit to the optical data alone results in a similar value, $\beta\approx0.7$, but with very large errors. Assuming an SMC--like dust density, we derive an extinction $A_V=0.24\pm0.06$ (note that this error is somewhat underestimated due to the constrained spectral slope). Alternatively, without dust, the optical data yield a red spectral slope of $\beta=1.41\pm0.18$. Using this extinction value and the redshift, we find a magnitude correction~\citep{Kann2006ApJ} of $dRc ={-3.46}^{+0.17}_{-0.23}$ (alternatively, $dRc =-3.09\pm0.13$ without extinction). Using the known SED, we are able to create a compound $R$-band light curve, to which we add GCN data~\citep{Malesani2011GCN,gcn12225}. The resulting light curve spans from 32 seconds to $\approx1.5$ days after the GRB, and reaches a peak at $9^{\rm{th}}$ magnitude, implying that this is initially an exceptionally bright afterglow.  Comparing the known sample of optical afterglows with prompt--emission detections in the high--MeV/GeV range~\citep[including one \emph{AGILE} GRID GRB, GRB~080514B,][]{Giuliani2008A&A,Rossi2008A&A,McBreen2010A&A,Cenko2011ApJ} we find that while some afterglows are among the most luminous known (GRB~090923, GRB~090926A), the others are of average luminosity at $t=1$ day, including \GRB. The span is four magnitudes, and if one extrapolates the steep late decay of the short GRB~090510~\citep{NicuesaGuelbenzu2012A&A}, the span reaches about ten magnitudes. The very early detection of the afterglow of \GRB also allows us for the first time to compare the brightness at 0.001 days (normalized at $z=1$) for a LAT--detected GRB, we find $R=10.5$.  Coupled with the unbroken decay from the earliest detection on, this makes it one of the brightest forward--shock dominated afterglows known~\citep{Kann2010ApJ}.

Our broadband spectral and temporal modeling, including LAT data, favors a circum--burst medium with a wind--like density profile and an afterglow onset time $\lesssim 9$~s after trigger.  This is the first time that a LAT--detected long GRB could have such an early onset time estimated using multiwavelength data.  The corresponding bulk Lorentz factor value $\Gamma_0 \sim 500$, deduced from the deceleration time scale of the fireball, is also better motivated for \GRB.  With a constant density ISM environment and by assuming that the deceleration time scale coincides with the peak LAT flux time at $0.7$~s, the initial bulk Lorentz factor of the short GRB~090510 was calculated to be $\Gamma_0 \sim 2000$~\citep{2010ApJ...709L.146D} .  Using $\Gamma_0$ above the lower limit $\Gamma_{\rm min}$, calculated from $\gamma\gamma$--opacity argument in the simple one--zone model in the prompt phase, \citet{Cenko2011ApJ} have modeled the temporally extended X--ray--optical--radio afterglow data of  4 LAT--detected long GRBs:  GRB~090323, GRB~090328, GRB~090902B and GRB~090926A.   A wind environment was preferred for each of these GRBs except for GRB~090902B, for which an ISM environment was preferred~\citep{Cenko2011ApJ}.  Although the sample is rather small, it hints at an interesting trend since a systematic study of well--sampled  long GRB afterglows shows that the majority of them  is consistent with a uniform density environment~\citep{2011A&A...526A..23S}.  The micro--physical parameters, $\epsilon_e$ and $\epsilon_B$ that we estimated for \GRB in Section~\ref{sec:afterglowparam}, are lower than for GRB~090510 with $\epsilon_e \sim \epsilon_B \sim 0.1$ which were also estimated from broadband spectral and temporal modeling including LAT data~\citep{2010ApJ...709L.146D}.  Values of $\epsilon_e$ and $\epsilon_B$ closer to our estimates have been obtained for GRB~090323 with late afterglow data \citep{Cenko2011ApJ}.


\section{Conclusions}
\label{sec:conclusions}
\GRB is the first long burst detected by instruments on--board \Fermi\ and \Swift\ with simultaneous coverage from optical to gamma rays for a few hundred seconds.  Prompt follow up observations by MOA and late--time data from GROND provide crucial optical data from a few minutes to a few days after the trigger and while the afterglow is still bright in the XRT.  The redshift of the burst $z=2.83$ is relatively high but within the range of other bright long bursts detected by \Fermi\ LAT.

\GRB has a rather high total energy output, with an isotropic--equivalent gamma--ray (10 keV--10 GeV) energy release of $E_{\rm iso} \sim 6\times 10^{53}$~erg within the prompt phase of the first $\sim 8.5$~s. The brightest peak in the LAT light curve at \Tz{5.5} coincides with peaks in all other energy bands, producing a peak isotropic--equivalent $\gamma$-ray (10 keV--10 GeV) luminosity of $\sim 6\times 10^{53}$~erg~s$^{-1}$.  The LAT emission is delayed by $\sim 2.5$~s from the GBM trigger and continues for the next $\sim 850$~s.  These features (delayed onset and temporally extended emission in the LAT) are common to many bright LAT bursts.  Similarly to other bright LAT bursts, we detect an additional power law component to the Band function in the time--integrated spectral analysis, with evidence also in the time--resolved analysis, although with lower significance. For \GRB, we also have  evidence for a cutoff in the power law component at GeV energies, although with a significance which is not as high as that measured for GRB~090926A. Indeed, two Comptonized models provide a slightly better fit to the time--integrated spectra from $T_{05,\rm{LAT}}$ to $T_{95,\rm{GBM}}$ with respect to the Band plus power law model. Using a measured variability time scale in the GBM data and by assuming that the LAT photons are co--spatial with softer photons, we calculate a jet bulk Lorentz factor of $\sim 500$ from the $\gamma\gamma$ attenuation mechanism producing the GeV spectral cutoff.  A value smaller by a factor $\sim 2$ to 3 is also acceptable within modeling uncertainties; for instance, we found $\sim 300$ using a  parametrization for two--zone emission models (c.f. Sec.~\ref{sec:disc:prompt}).

The broadband spectrum of \GRB from optical to $\gtrsim 100$~MeV at \Tz{550}, when fit with a broken power law, together with temporal flux decay behavior in different wavelengths, favors an afterglow wind model as the origin.  We interpret the spectral steepening as a cooling break and verify its expected temporal evolution by fitting the broadband spectrum at 2.74~d.  The presence of two mild X--ray flares and lack of good optical data at an earlier epoch (III) prohibit us from independently verifying the afterglow model.  However, the broadband emission, after extraction of the X--ray flares, is consistent with the model at \Tz{550}.  Most remarkably, we find that BAT and LAT spectra at $9$~s and $16$~s can be fit with single power laws with compatible spectral indices, which in turn are consistent with the measured X--ray spectral index at later times.  These broadband spectra at early times strongly suggest their origin as afterglow emission as well.  This scenario, however, requires an afterglow onset time before $\sim 8$~s.

With an afterglow onset time of $5.9$~s, first time interval after the brightest LAT peak, we interpret subsequent multiwavelength data as originating from forward shock emission in a wind--type medium.  We cannot rule out contamination starting at an early time from prompt emission, however.  With an initial bulk Lorentz factor $\sim$ 500--550 and wind density parameter within the acceptable range we are able to fit and interpret the broadband data. It is worth emphasizing that this second derivation of the bulk Lorentz factor, using the afterglow modeling, is independent from the estimate done using the cutoff during the prompt phase, but that both numbers are nevertheless very compatible. The resulting micro--physical parameters are $\epsilon_e \sim 3\times 10^{-3}$ and $\epsilon_B \sim 10^{-2}$ which favor a somewhat larger magnetic energy density than electron energy density in the forward shock.  The time scale for the fast--to--slow cooling spectral transition is also compatible with earlier single power law spectra and later broken power law spectra.  The total isotropic--equivalent jet kinetic energy calculated is $\sim 3\times 10^{54}$~erg which puts \GRB among the most energetic bursts.

Compared to other LAT GRBs, \GRB is in the middle of the fluence distribution and is not as bright as the four brightest GRBs, but due to favorable observing conditions we were able to collect comprehensive multiwavelength data and put strong constraints on the temporally extended GeV emission. In addition, both X--ray and optical data strongly suggest the presence of early forward--shock emission in this burst, which would constitute the earliest afterglow emission detected so far among LAT--detected GRBs.

We find that temporally extended LAT emission is compatible with originating from the forward shock that produces the broadband afterglow emission. In the context of this scenario, the delayed onset of the LAT emission can be interpreted as the increasing flux of the forward--shock component, before reaching the deceleration time and after which the traditional afterglow phase begins.  The validity of this scenario and the origin of the LAT emission will be further tested by future multiwavelength observations of \Fermi LAT--detected bursts.

\acknowledgments
The \Fermi\ GBM collaboration acknowledges support for GBM development, operations and data analysis from NASA in the US and BMWi/DLR in Germany.
The \Fermi LAT Collaboration acknowledges generous ongoing support from a number of agencies and institutes that have supported both the development and the operation of the LAT as well as scientific data analysis. These include the National Aeronautics and Space Administration and the Department of Energy in the United States, the Commissariat \`a l'Energie Atomique and the Centre National de la Recherche Scientifique / Institut National de Physique Nucl\'eaire et de Physique des Particules in France, the Agenzia Spaziale Italiana and the Istituto Nazionale di Fisica Nucleare in Italy, the Ministry of Education, Culture, Sports, Science and Technology (MEXT), High Energy Accelerator Research Organization (KEK) and Japan Aerospace Exploration Agency (JAXA) in Japan, and the K.~A.~Wallenberg Foundation, the Swedish Research Council and the Swedish National Space Board in Sweden. Additional support for science analysis during the operations phase is gratefully acknowledged from the Istituto Nazionale di Astrofisica in Italy and the Centre National d'\'Etudes Spatiales in France.

We gratefully acknowledge the contributions of dozens of members of the
Swift team at OAB, PSU, UL, GSFC, ASDC, and MSSL and our subcontractors,
who helped make these instruments possible. This work made use of data supplied by the UK \Swift Science Data Centre at the University of Leicester. S.R.O acknowledges support from the UK Space Agency.

Part of the funding for GROND (both hardware as well as personnel) was generously granted from the Leibniz-Prize to Prof. G. Hasinger (DFG grant HA 1850/28-1). TK acknowledges support by the European Commission under the Marie Curie Intra-European Fellowship Programme. D.A.K. and S.K. acknowledge support by grant DFG Kl 766/16-1. D.A.K. is grateful for travel funding support through the MPE.

We would like to acknowledge the MOA collaboration to permit target of opportunity observations for GRB afterglow. We also would like to acknowledge the University of Canterbury for allowing MOA to use the B\&C telescope. This work was partially supported by the Ministry of Education, Culture, Sports, Science and Technology (MEXT) of Japan.

\newpage
\bibliographystyle{apj}
\bibliography{master,extra}       

\newpage
\begin{appendix}
\section{Data tables}

\begin{table}[htbp] 
\centering 
\begin{tabular}{lccc} 
\hline\hline 
Time     & Energy & Photon Flux above 100 MeV & Test Statistic\\ 
Bins (s) & Index  & (ph cm$^{-2}$s$^{-1}$) & \\\hline\hline
(1) 2.35--3.59 & $-2.22\pm0.39$ & $1.11\pm0.40\,\times10^{-3}$      &  73  \\
(2) 3.59--4.56 & $-2.83\pm0.60$ & $1.47\pm0.55\,\times10^{-3}$      &  53  \\
(3) 4.56--5.47 & $-2.45\pm0.49$ & $1.45\pm0.57\,\times10^{-3}$      &  68  \\
(4) 5.47--5.67 & $-2.25\pm0.41$ & $6.80\pm2.48\,\times10^{-3}$      &  88  \\
(5) 5.67--6.58 & $-3.12\pm0.68$ & $1.69\pm0.60\,\times10^{-3}$      &  90  \\
(6) 6.58--8.27 & $-2.11\pm0.37$ & $7.66\pm2.87\,\times10^{-4}$      &  70  \\
(7) 8.27--11.54 & $-1.80\pm0.31$ & $3.50\pm1.42\,\times10^{-4}$     &  73  \\
(8) 11.54--19.73 & $-4.65\pm1.24$ & $1.92\pm0.71\,\times10^{-4}$    &  41  \\
(9) 19.73--32.60 & $-3.13\pm0.82$ & $8.39\pm3.68\,\times10^{-5}$    &  27  \\
(10) 32.60--110.97 & $-1.92\pm0.38$ & $1.24\pm0.65\,\times10^{-5}$  &  20  \\
(11) 110.97--227.04 & $-3.01\pm0.68$ & $1.64\pm0.71\,\times10^{-5}$ &  17  \\
(12) 227.04--853.89 & $-1.69\pm0.35$ & $1.77\pm1.16\,\times10^{-6}$ &  16  \\
(13) 853.89--1433.65 & $-2.25$ & $<3.9\,\times10^{-6}$              &   2  \\\hline\hline
 \end{tabular} 
\caption{LAT Time--resolved spectroscopy data, photon fluxes and photon indices. The last entry is an upper--limit assuming a power--law index of $-2.25$.} 
\label{Tab:LATExt} 
\end{table} 

\begin{table}[htbp]
\begin{center}
\scalebox{0.8}{%
\begin{tabular}{ccrrccccc}
\hline\hline
Time (s)&   Exposure(s) &  Magnitude &     Flux (mJy) & Filter \\
\hline\hline
62	& 9	   & $ 13.08^{+0.06}_{-0.06}$  & $21.26\pm0.71  $ &  $v$          \\
627	& 20	   & $ 16.37^{+0.28}_{-0.22}$  & $ 1.03\pm0.14  $ &  $v$          \\
800	& 20	   & $ 16.86^{+0.43}_{-0.30}$  & $ 0.66\pm0.13  $ &  $v$          \\
1100	& 619	   & $ 17.48^{+0.34}_{-0.26}$  & $ 0.37\pm0.06  $ &  $v$          \\
12753	& 436	   & $ 20.30^{+3.36}_{-0.73}$  & $ 0.02\pm0.06  $ &  $v$          \\
1557278	& 1069878  & $>20.42                $  & $<0.03         $ &  $v$          \\
552 	& 20	   & $ 17.62^{+0.40}_{-0.29}$  & $ 0.36\pm0.06  $ &  $b$          \\
725	& 20	   & $ 18.02^{+0.59}_{-0.38}$  & $ 0.25\pm0.05  $ &  $b$          \\
1239	& 192	   & $ 18.91^{+0.68}_{-0.41}$  & $ 0.11\pm0.03  $ &  $b$          \\
15599	& 17857	   & $ 21.15^{+3.54}_{-0.73}$  & $ 0.01\pm0.01  $ &  $b$          \\
1557674	& 1070041  & $>21.00                $  & $<0.02         $ &  $b$          \\
299	& 25	   & $ 17.28^{+0.33}_{-0.25}$  & $ 0.18\pm0.02  $ &  $u$          \\
324	& 25	   & $ 17.05^{+0.28}_{-0.22}$  & $ 0.22\pm0.02  $ &  $u$          \\
349	& 25	   & $ 16.98^{+0.26}_{-0.21}$  & $ 0.23\pm0.02  $ &  $u$          \\
374	& 25	   & $ 16.76^{+0.23}_{-0.19}$  & $ 0.28\pm0.02  $ &  $u$          \\
399	& 25	   & $ 17.11^{+0.29}_{-0.23}$  & $ 0.21\pm0.02  $ &  $u$          \\
424	& 25	   & $ 17.53^{+0.39}_{-0.29}$  & $ 0.14\pm0.02  $ &  $u$          \\
449	& 25	   & $ 17.57^{+0.41}_{-0.29}$  & $ 0.13\pm0.02  $ &  $u$          \\
474	& 25	   & $ 17.59^{+0.41}_{-0.29}$  & $ 0.13\pm0.02  $ &  $u$          \\
499	& 25	   & $ 17.35^{+0.34}_{-0.26}$  & $ 0.16\pm0.02  $ &  $u$          \\
524	& 25	   & $ 18.08^{+0.62}_{-0.39}$  & $ 0.08\pm0.02  $ &  $u$          \\
1001	& 620	   & $>18.23                $  & $ 0.01\pm0.01  $ &  $u$          \\
21275	& 29621	   & $>20.37                $  & $<0.01         $ &  $u$          \\
439746	& 415954   & $>21.1                 $  & $<0.01         $ &  $u$          \\
976	& 620	   & $>18.14                $  & $<0.05         $ &  $uvw1$       \\
951	& 620	   & $>19.12                $  & $<0.02         $ &  $uvm2$       \\
989	& 793      & $>18.1                 $  & $<0.04         $ &  $uvw2$ \\\hline\hline
\end{tabular}
}
\end{center}
\caption{\Swift\ UVOT data table for the individual filters. Time is the mid--time in exposure, in seconds, since BAT trigger. Exposure is the half--width of the integration duration in seconds. Magnitudes and flux densities have been corrected for Galactic extinction only. Upper limits are given at $3~\sigma$ for both magnitudes and fluxes.}
\label{Tab:uvot:uvb}
\end{table}

\begin{table}[htbp]
\begin{center}
\scalebox{0.8}{%
\begin{tabular}{ccrrccccc}
\hline\hline
Time (s)&   Expo(s) &  Magnitude &     Flux (mJy) & Filter \\
\hline\hline
80	& 10	   & $14.47^{+0.07}_{-0.06}$  & $3.13\pm0.08$ &  $white$      \\
90	& 10	   & $14.55^{+0.07}_{-0.06}$  & $2.91\pm0.08$ &  $white$      \\
100	& 10	   & $14.68^{+0.07}_{-0.07}$  & $2.59\pm0.07$ &  $white$      \\
110	& 10	   & $14.87^{+0.07}_{-0.07}$  & $2.17\pm0.06$ &  $white$      \\
120	& 10	   & $15.05^{+0.08}_{-0.07}$  & $1.83\pm0.06$ &  $white$      \\
130	& 10	   & $15.08^{+0.08}_{-0.07}$  & $1.79\pm0.05$ &  $white$      \\
140	& 10	   & $15.11^{+0.08}_{-0.07}$  & $1.74\pm0.05$ &  $white$      \\
150	& 10	   & $15.29^{+0.08}_{-0.08}$  & $1.48\pm0.05$ &  $white$      \\
160	& 10	   & $15.36^{+0.08}_{-0.08}$  & $1.38\pm0.05$ &  $white$      \\
170	& 10	   & $15.52^{+0.09}_{-0.08}$  & $1.19\pm0.04$ &  $white$      \\
180	& 10	   & $15.49^{+0.09}_{-0.08}$  & $1.22\pm0.04$ &  $white$      \\
190	& 10	   & $15.6 ^{+0.09}_{-0.08}$  & $1.11\pm0.04$ &  $white$      \\
200	& 10	   & $15.73^{+0.10}_{-0.09}$  & $0.98\pm0.04$ &  $white$      \\
210	& 10	   & $15.83^{+0.10}_{-0.09}$  & $0.89\pm0.03$ &  $white$      \\
220	& 10	   & $15.91^{+0.10}_{-0.09}$  & $0.83\pm0.03$ &  $white$      \\
577	& 20	   & $17.12^{+0.16}_{-0.14}$  & $0.27\pm0.02$ &  $white$      \\
750	& 20	   & $17.53^{+0.22}_{-0.18}$  & $0.19\pm0.02$ &  $white$      \\
940	& 150	   & $18.26^{+0.14}_{-0.13}$  & $0.10\pm0.01$ &  $white$      \\
1178	& 20	   & $19.32^{+1.93}_{-0.66}$  & $0.04\pm0.01$ &  $white$      \\
1350	& 20	   & $18.48^{+0.55}_{-0.36}$  & $0.08\pm0.01$ &  $white$      \\
6975	& 200	   & $20.27^{+0.73}_{-0.43}$  & $0.01\pm0.01$ &  $white$      \\
104187	& 14049	   & $>20.96               $  & $<0.01      $ &  $white$   \\\hline\hline
\end{tabular}
}
\end{center}
\caption{\Swift\ UVOT data table for the co-added exposures.  Time is the mid--time in exposure, in seconds, since GBM trigger. Exposure is the half--width of the integration duration in seconds. Magnitudes and flux densities have been corrected for Galactic extinction only. Upper limits are given at $3~\sigma$ for both magnitudes and fluxes.}
\label{Tab:uvot:white}
\end{table}


\begin{table}[htbp]
\begin{center}
\scalebox{0.8}{%
\begin{tabular}{cccc||cccc}
\hline\hline
Time (s)   &  Magnitude $\dagger$ & Flux (mJy)$\ddagger$ & Filter &Time (s) & Magnitude $\dagger$ & Flux (mJy)$\ddagger$ & Filter  \\ \hline \hline
       205    &  14.50    $\pm$   0.10   & $4.160^{+0.408}_{-0.372}$ & $I$  &  2060  &    18.15    $\pm$   0.19   & $0.144^{+0.028}_{-0.023}$ & $I$\\
       268    &  14.87    $\pm$   0.10   & $2.960^{+0.293}_{-0.266}$ & $I$  &  2143  &     19.15   $\pm$     0.20 & $0.082^{+0.017}_{-0.014}$ & $V$\\
       333    &  15.22    $\pm$   0.10   & $2.160^{+0.218}_{-0.198}$ & $I$  &  2208  &     19.23   $\pm$     0.20 & $0.076^{+0.016}_{-0.013}$ & $V$\\
       428    &   16.76   $\pm$   0.10   & $0.746^{+0.071}_{-0.065}$ & $V$  &  2271  &     19.33   $\pm$     0.22 & $0.070^{+0.015}_{-0.013}$ & $V$\\
       556    &   17.19   $\pm$   0.10   & $0.503^{+0.051}_{-0.046}$ & $V$  &  2362  &    18.33    $\pm$   0.20   & $0.122^{+0.025}_{-0.021}$ & $I$\\
       656    &  16.29    $\pm$   0.11   & $0.800^{+0.089}_{-0.080}$ & $I$  &  2426  &    18.14    $\pm$   0.19   & $0.146^{+0.028}_{-0.023}$ & $I$\\
       720    &  16.42    $\pm$   0.12   & $0.709^{+0.080}_{-0.072}$ & $I$  &  2489  &    18.37    $\pm$   0.20   & $0.119^{+0.024}_{-0.020}$ & $I$\\
       785    &  16.54    $\pm$   0.12   & $0.639^{+0.074}_{-0.066}$ & $I$  &  2573  &     19.47   $\pm$     0.22 & $0.061^{+0.014}_{-0.011}$ & $V$\\
       868    &   17.88   $\pm$   0.12   & $0.266^{+0.031}_{-0.028}$ & $V$  &  2636  &     19.44   $\pm$     0.23 & $0.063^{+0.015}_{-0.012}$ & $V$\\
       933    &   17.97   $\pm$   0.12   & $0.244^{+0.029}_{-0.026}$ & $V$  &  2787  &    18.59    $\pm$   0.22   & $0.096^{+0.022}_{-0.018}$ & $I$\\
       996    &   17.99   $\pm$   0.12   & $0.2410^{+0.029}_{-0.026}$ & $V$ &  2851  &    18.51    $\pm$   0.22   & $0.104^{+0.023}_{-0.019}$ & $I$\\
      1082    &  17.09    $\pm$   0.13   & $0.384^{+0.050}_{-0.045}$ & $I$  &  2914  &    18.53    $\pm$   0.22   & $0.102^{+0.023}_{-0.019}$ & $I$\\
      1145    &  17.16    $\pm$   0.14   & $0.359^{+0.049}_{-0.043}$ & $I$  &  2999  &     19.87   $\pm$     0.28 & $0.042^{+0.013}_{-0.096}$ & $V$\\
      1210    &  17.30    $\pm$   0.14   & $0.315^{+0.044}_{-0.039}$ & $I$  &  3127  &     19.77   $\pm$     0.27 & $0.047^{+0.013}_{-0.010}$ & $V$\\
      1293    &   18.52   $\pm$   0.14   & $0.148^{+0.021}_{-0.018}$ & $V$  &  3214  &    18.59    $\pm$   0.23   & $0.097^{+0.023}_{-0.018}$ & $I$\\
      1358    &   18.55   $\pm$   0.15   & $0.143^{+0.021}_{-0.018}$ & $V$  &  3277  &    18.74    $\pm$   0.25   & $0.084^{+0.022}_{-0.017}$ & $I$\\
      1421    &   18.53   $\pm$   0.15   & $0.146^{+0.021}_{-0.018}$ & $V$  &  3342  &    18.59    $\pm$   0.23   & $0.096^{+0.023}_{-0.018}$ & $I$\\
      1506    &  17.68    $\pm$   0.16   & $0.222^{+0.035}_{-0.031}$ & $I$  &  3637  &    19.00    $\pm$   0.29   & $0.066^{+0.020}_{-0.015}$ & $I$\\
      1569    &  17.81    $\pm$   0.17   & $0.198^{+0.033}_{-0.029}$ & $I$  &  3978  &     19.73   $\pm$     0.25 & $0.048^{+0.013}_{-0.010}$& $V$\\
      1634    &  17.70    $\pm$   0.16   & $0.219^{+0.035}_{-0.030}$ & $I$  &  4061  &    18.71    $\pm$   0.24   & $0.086^{+0.0211}_{-0.017}$ & $I$\\
      1717    &   18.84   $\pm$   0.17   & $0.110^{+0.019}_{-0.016}$ & $V$  &  4126  &    18.97    $\pm$   0.28   & $0.068^{+0.020}_{-0.015}$ & $I$\\
      1783    &   18.97   $\pm$   0.18   & $0.098^{+0.018}_{-0.015}$ & $V$  &  4188  &    18.99    $\pm$   0.28   & $0.067^{+0.020}_{-0.015}$ & $I$\\
      1846    &   18.89   $\pm$   0.17   & $0.105^{+0.018}_{-0.016}$ & $V$  &  4485  &    19.16    $\pm$   0.30   & $0.057^{+0.018}_{-0.014}$ & $I$\\
      1931    &  18.03    $\pm$   0.18   & $0.162^{+0.030}_{-0.025}$ & $I$  &  4548  &    19.02    $\pm$   0.28   & $0.065^{+0.019}_{-0.015}$ & $I$\\
      1994    &  18.03    $\pm$   0.19   & $0.162^{+0.031}_{-0.026}$ & $I$  &  -     &    -                       & -                         & -\\\hline
\end{tabular}
}
\end{center}
$\dagger$ Extinction correction: $A_{\lambda}^{V}$ = 0.635, $A_{\lambda}^{I}$ = 0.371
$\ddagger$ Flux density
\caption{MOA data table used in the $V$ and $I$ bands. Time is given since the GBM trigger. The exposure for each frame is 30~s.}
\label{Tab:MOA-all}
\end{table}

\begin{table}[htbp]
\begin{center}
\begin{tabular}{ccrrccccc}
\hline\hline
Days   & Filter & Mag (Vega) & Mag (AB) & Flux ($\mu$Jy)$\dagger$
& Ext. (Mag)$\ddagger$ & Exp. time (s)\\ \hline \hline
2.7404 &  $g^\prime$  & $25.31 \pm0.24$ & $25.25\pm0.24$  &  $0.54^{+0.13}_{-0.11}$ & 0.68 &  4500     \\ 
2.7404 &  $r^\prime$  & $23.66 \pm0.10$ & $23.84\pm0.10$  &  $1.63^{+0.15}_{-0.14}$ & 0.47 &  4500     \\ 
2.7404 &  $i^\prime$  & $23.19 \pm0.11$ & $23.60\pm0.11$  &  $1.82^{+0.20}_{-0.18}$ & 0.35 &  4500     \\ 
2.7404 &  $z^\prime$  & $22.59 \pm0.10$ & $23.13\pm0.10$  &  $2.58^{+0.25}_{-0.23}$ & 0.26 &  4500     \\ 
2.7404 &  $J$   & $21.66 \pm0.29$ & $22.59\pm0.29$  &  $3.79^{+1.15}_{-0.88}$ & 0.15 &  3600     \\ 
2.7404 &  $H$   & $20.92 \pm0.28$ & $22.31\pm0.28$  &  $4.67^{+1.40}_{-1.08}$ & 0.10 &  3600     \\ 
2.7404 &  $K$   & $> 18.2       $ & $> 20.1      $  &  $< 35.5              $ & 0.06 &  3600     \\ \hline 
\end{tabular}
\end{center}
$\dagger$ Flux density, corrected for Galactic extinction.\\
$\ddagger$ The Galactic extinction correction along the line of sight for $E(B-V)=0.175$ \citep{SFD1998} using a CCM Milky-Way extinction law \citep{CCM1989}.
\caption{GROND Data table; seeing during the observation ranged from 1\farcs2 to 1\farcs5 depending on the band.}
\label{Tab:grond}
\end{table}

\end{appendix}

\end{document}